\documentclass[12pt]{article}
\usepackage{amssymb}
\usepackage{graphicx,epstopdf}
\usepackage{amsthm}
\usepackage{tikz}
\usepackage{subfig}
\usepackage[authoryear, longnamesfirst]{natbib}
\usepackage[lined,algonl,boxed]{algorithm2e}
\usepackage{multicol}
\usepackage{paralist}
\usepackage{url}
\usepackage{amsmath}
\usepackage{amsfonts}
\usepackage{color}
\usepackage{afterpage}

\newcommand{\field}[1]{\mathbb{#1}}
\newcommand{\R}{\field{R}}

\newcommand{\p}{\field{P}}
\newcommand{\N}{\field{N}}
\newcommand{\Z}{\field{Z}}

\newcommand{\E}{\field{E}}

\newcommand{\BL}{\text{BL}}
\newcommand{\PP}{\mathbb{P}}
\newcommand{\GG}{\mathbb{G}}
\newcommand{\pstar}{\stackrel{\PP^*}{\rightarrow}}

\newcommand{\Ac}{\mathcal{A}}
\newcommand{\Fc}{\mathcal{F}}
\newcommand{\Gc}{\mathcal{G}}
\newcommand{\Hc}{\mathcal{H}}

\newcommand{\Xc}{\mathcal{X}}

\newcommand\weakP[1]{\ \overset{\PP}{\underset{#1}{\rightsquigarrow}}\ }
\newcommand\weak{\rightsquigarrow}

\newcommand{\Var}{\mathrm{Var}}
\newcommand{\eps}{\varepsilon}

\theoremstyle{example}
\theoremstyle{remark}
\theoremstyle{lemma}
\theoremstyle{definition}
\theoremstyle{corol}
\theoremstyle{proposition}
\theoremstyle{condition}
\theoremstyle{assumption}

\newtheorem{theorem}{\n{Theorem}}[section]

\newtheorem{remark}{\n{Remark}}[section]

\font\n=cmcsc10
\def\cov{{\mbox{cov}}}

\begin{document}

\title{{\Large\bf A subsampled double bootstrap for massive data} }
\author{Srijan Sengupta, Stanislav Volgushev, and Xiaofeng Shao\thanks{Srijan Sengupta (ssengpt2@illinois.edu) is Ph.D. candidate, Department of
Statistics, University of Illinois at Urbana-Champaign. Stanislav Volgushev (stanislav.volgushev@rub.de) is postdoctoral associate at Ruhr-Universit\"at Bochum. Xiaofeng Shao (xshao@illinois.edu) is Associate Professor, Department of Statistics, University of Illinois at Urbana-Champaign. Stanislav Volgushev is partially supported by the SFB 823 (Teilprojekt C1) of the German research foundation (DFG). Shao acknowledges partial financial support from National Science Foundation grants  DMS-110454 and DMS-1407037. Xiaofeng Shao would like to thank Guang Cheng
for bringing the work of Kleiner et al. (2014) to his attention. 
We thank the Editor and two anonymous referees for their helpful comments.
}}
\maketitle

{Abstract: The bootstrap is a popular and powerful method for {assessing} 
precision of estimators and inferential methods. However, for massive datasets which are increasingly prevalent, the bootstrap becomes prohibitively costly in computation and its feasibility is questionable even with modern parallel computing platforms. Recently \cite{kleiner2014scalable} proposed a method called BLB (Bag of Little Bootstraps) for massive data which is more computationally scalable with little sacrifice of statistical accuracy. Building on BLB and the idea of fast double bootstrap, we propose a new resampling method, the subsampled double bootstrap, for both independent data and time series data. {We establish consistency of the subsampled double bootstrap under mild conditions for both independent and dependent cases.} Methodologically, the subsampled double bootstrap is superior to BLB in terms of running time, more sample coverage and automatic implementation with less tuning parameters for a given time budget. Its advantage relative to BLB and bootstrap is also demonstrated in numerical simulations and a data illustration.
}\\

Keywords: Big Data, Computational cost, Subsampling, Resampling
\section{Introduction}

In the past decade, we have witnessed massive data (or big data) generated in many fields.
Datasets grow in size in part because they are increasingly being collected by ubiquitous information-sensing mobile devices, remote sensing technologies, and wireless sensor networks, among others.
Although our computing power has also been advancing steadily, the surge of massive data presents challenges to both computer scientists and statisticians in terms of data storage, computation and statistical analysis.
As nicely summarized in \cite{jordan2013statistics}, a key question for statistical inference in the massive data context is ``Can you guarantee a certain level of inferential accuracy within a certain time budget even as the data grow in size''?
{From a statistical point of view, there is a great need for new methods that are theoretically sound and remain computationally feasible even for massive data sets.}
The classical theoretical criteria to assess the quality of an inferential procedure such as mean squared error, size/power are still relevant, but for massive data, computational efficiency and algorithm quality are also important considerations in comparing different statistical methods and procedures.

With any statistical inference method, an inextricably associated problem is to assess the precision of that inference, and this remains important for the statistical analysis of massive data sets.
For example after parameter estimation from a data set, a natural next step  is to measure how precise the estimation method is, and this can be measured by the mean squared error, width of confidence interval, and so on.
The bootstrap (\cite{efron1979bootstrap}) is a powerful and popular procedure that can be applied to estimate precision for a wide variety of inference methods.
It has well-known statistical properties including consistency and higher-order accuracy under quite general settings.
It is conceptually appealing as it is straightforward to implement, using resamples from the data as a proxy for samples, and is automatic in nature {such that} the user can implement it without advanced statistical knowledge.
However, the benefits of bootstrap come at a considerable computational cost.
Each iteration of the bootstrap involves the calculation of a statistical function on a resample of the original data.
For a data of size $n$, on average each resample includes 0.63$n$ distinct sample points --- therefore each iteration of the bootstrap carries a computational cost of the same order as that of the original inference on the data.
Even though this problem can be alleviated with the advent of modern parallel computing platforms, it is still quite overwhelming to repeatedly process such resampled datasets for data of huge size, say a terabyte.
Therefore, this calls for new bootstrap methods that are computationally scalable while maintaining good statistical properties.

In their recent work \citet*{kleiner2014scalable} introduced a new resampling procedure called Bag of Little Bootstraps (BLB, hereafter).
This procedure consists of randomly selecting small subsets of the data, and then performing a bootstrap on each subset, by constructing weighted  resamples of the subset such that the resample size equals the size of the original data.
The estimator is calculated on these resamples in the same manner as bootstrap.
It is worth noting that this method bears some resemblance to the traditional  subsampling (\cite{politis1994large}) or $m$ out of $n$ bootstrap (\cite{bickel1997resampling}), which  involve subsamples or resamples of size much smaller than the bootstrap, thereby reducing the computational cost.
However, these methods (subsampling or $m$ out of $n$ bootstrap) usually require a rescaling of the output, to adjust for the difference between sample size and resample or subsample size. 
This feature makes them less user-friendly, since in order to evaluate the precision of an estimation method, the practitioner typically needs to know the rate of convergence of the estimator being used. {Additionally, as demonstrated in \citet{kleiner2014scalable}, the performance of subsampling or $m$ out of $n$ bootstrap depends quite strongly on the choice of parameters such as subsample size.} By contrast, the resamples in BLB are of the same size as the data, so no rescaling of output is needed
 thereby retaining the automatic and user-friendly nature of the bootstrap.
On the other hand, although the resamples are nominally of the same size as the original data, they contain only a small number of distinct points coming from the subset, which reduces the computational cost of calculating a statistical function of the resamples.
The estimates of precision from a few subsets can be averaged to obtain the BLB estimate of precision.

In \cite{kleiner2014scalable} the authors recommend a large number of resamples from each subset, and a small number of random subsets.
However, this means that only a small fraction of the original data is used in computing the BLB estimate, as a large majority of data points may not appear in any of the subsets used.
Additionally, running a large number of resamples on each subset might incur high computational costs, even if each resample has less runtime than bootstrap resamples.
These two issues can affect the performance of BLB in terms of statistical
accuracy and computational cost, respectively.

When facing the trade-off between statistical accuracy and computational cost, a practical question we need to answer is:
``given a certain computation time budget, how can a practitioner optimally use that budget to come up with an estimate of precision?''
The bootstrap has an obvious answer to this question --- keep taking resamples until the budget runs out.
This answer holds true irrespective of the statistical inferential method whose precision is of interest.
However, it is not obvious how to answer this question for BLB, since it is not clear how to optimize the two tuning parameters --- namely number of resamples per subset and the number of subsets, under the time budget constraint.
Two natural strategies would be --- with a fixed number of resamples per subset use as many random subsets as possible, or with a fixed number of random subsets use as many resamples per subset as possible.
Both strategies might be sub-optimal in practice, depending upon the particular problem at hand. 
\cite{kleiner2014scalable} suggest a novel adaptive method for selecting the tuning parameters, where one first fixes a tolerance parameter, and then for each subset, one can keep taking resamples till that tolerance level is reached.
This method provides a nice way of adaptively choosing tuning parameters for a given level of desired accuracy.
However for a given computational time budget, the variability of the precision estimate is not known a priori,  and hence it is not clear how to choose an appropriate value of the tolerance parameter that is neither too ambitious nor too conservative for the inference method of interest.


In this article we present a new resampling procedure called the Subsampled Double Bootstrap (SDB, hereafter) for massive data.
Double bootstrap was first proposed by \cite{beran1988prepivoting} as a way of improving the accuracy of bootstrap, but is considerably more expensive than bootstrap and becomes computationally infeasible for massive data.
Fast double bootstrap (FDB, hereafter), which was independently proposed by \cite{white2000reality} and Davidson and MacKinnon (\citeyear{davidson2000improving},\citeyear{davidson2007improving}), is an interesting alternative that only resamples once in the second stage of bootstrapping and can dramatically speed up the double bootstrap.
The FDB has been applied to many tests in econometrics, see \cite{davidson2002fast}, \cite{ahlgren2008bootstrap}, \cite{richard2009modified}, among others.
Recently, \cite{giacomini2013warp} applied the idea of FDB to reduce the computational cost in running Monte carlo experiments to assess the performance of bootstrap estimators and tests.
They demonstrated the consistency of this method and called it a `warp-speed method' to emphasize its rapidness.
\cite{chang2014double} recently studied the higher order accuracy of FDB in terms of bias correction and coverage accuracy of confidence intervals.
In the massive data context, the FDB is still too expensive since its computational cost is about twice the cost of bootstrap.
Therefore we propose to do subsampling first and then apply the idea of a single resample in the double bootstrap step to the randomly drawn subset of massive data, to evaluate the precision of a statistical inference method.
Since our method is a combination of subsampling and double bootstrap, we call it subsampled double bootstrap (SDB).
In the implementation of SDB, we randomly draw a large number of small subsets of the data, but instead of bootstrapping the subsets we construct only one resample from each subset.
Since these resamples have the same nominal size as the original data but only a small number of distinct points, SDB retains the automatic nature and computational strength of BLB.
The ensemble of resamples is then used to estimate the precision of the inference method, in the same manner as bootstrap.
Note that SDB inherits certain features from FDB but is computationally much cheaper than FDB.
The number of distinct points in the first-stage subsample and the second-stage resample of SDB are small compared to the number of distinct points in the first-stage and second-stage resamples of FDB, and this makes SDB much faster.

To see the statistical and computational advantages of SDB, note that the estimation time of one SDB iteration is comparable to that of two resamples for a BLB subset, and hence SDB can cover a large number of random subsets in the time it takes BLB to complete a large number of resamples for a single random subset.
Hence, SDB can provide a much more comprehensive coverage of the data than BLB {within a given time budget}.
Further, given a certain computational time budget, utilizing that budget with SDB is straightforward as it does not require the choice of any tuning parameters.
The practitioner can, just like bootstrap, simply keep running subset-resamples until the time budget runs out.

The {rest} of the article is organized as follows.
In Section \ref{sec:ind} we describe SDB in independent data setting.
Section \ref{sec:ind_theory} demonstrates the consistency of SDB for independent data, and Section \ref{sec:ind_sim} reports two simulation studies comparing SDB, BLB, and bootstrap for independent data.
We introduce a time series version of SDB in Section \ref{sec:ts}.
Section \ref{sec:ts_theory} demonstrates consistency for the dependent case, and Section \ref{sec:ts_sim} reports two simulation studies for time series data.
We provide a data illustration on a large meteorological time series dataset in Section \ref{sec:data}, and the article concludes with discussion in Section \ref{sec:disc}.
Proofs of the theoretical results are in the Appendix and some simulation figures are in supplementary materials.
The R codes used for simulation and data analysis are also in supplementary materials, as well as the dataset analyzed.

\section{SDB for independent data}
\label{sec:ind}
Consider an i.i.d. sample $\mathcal{X}_n = \{X_1,\ldots,X_n \}$ drawn from some unknown distribution $P$.
The parameter of interest is $\theta=\theta(P)$, for which an estimate $\hat{\theta}_n = \hat{\theta}(\mathcal{X}_n)$ is obtained from the sample.
(Please see the discussion following Theorem \ref{th:SDBiid} for a more rigorous definition of the types of parameter and estimator covered under the scope of SDB.)
Having chosen the estimator, the statistician often seeks to obtain further information regarding the precision of the estimator $\hat{\theta}(\mathcal{X}_n)$.
This requires the estimation of some measure involving the sampling distribution of $\hat{\theta}(\mathcal{X}_n)$ and the true value of the parameter $\theta$.
For example, the precision of an estimator can be measured by the mean squared error or the width of a 95\% confidence interval for $\theta$.

Such measures of precision can usually be defined in terms of a root function $T_n(\hat{\theta}_n, \theta)$ involving the estimator and the parameter.
Let $Q_n = Q_n(P)$ be the unknown sampling distribution of $T_n$, {and assume that the precision measure can be represented as $\xi(Q_n)$ for a suitable functional $\xi(\cdot)$.}
For example, suppose $\theta$ is the population mean, $\hat{\theta}(\mathcal{X}_n)$ is the sample mean, and the measure of interest is the scaled MSE $n E[(\hat{\theta}_n- \theta)^2]$.
In our notation, we define the root as $T_n(\hat{\theta}_n, \theta) = \sqrt{n}(\hat{\theta}_n- \theta)$, and define the functional as $\xi(Q_n) = \int x^2 dQ_n(x)$, where $Q_n$ is the sampling distribution of $T_n(\hat{\theta}_n, \theta)$.

Estimation of $\xi(Q_n)$ can be performed by a resampling method like bootstrap.
Let $\mathbb{P}_n$ be the empirical distribution of the sample $\mathcal{X}_n$, then we can approximate $Q_n(P)$ by $\hat{Q}_n = Q_n(\mathbb{P}_n)$.
To do so, we generate a large number ($R$) of resamples $\mathcal{X}^{*j}_n = \{X_{j_1},\ldots,X_{j_n} \}, j=1,\ldots,R$ from the observed sample $\mathcal{X}_n$.
Treating the original estimate $\hat{\theta}(\mathcal{X}_n)$ as the population parameter and the resample estimate $\hat{\theta}(\mathcal{X}^{*j}_n)$ as an estimated value of this parameter, we compute the root $T_n(\hat{\theta}(\mathcal{X}^{*j}_n), \hat{\theta}(\mathcal{X}_n))$ for each resample, and obtain the empirical distribution $\hat{Q}_{n,R}$ of this ensemble of roots.
Conditionally on $\mathcal{X}_n$, the empirical distribution $\hat{Q}_{n,R}$ converges to the resampling distribution $Q_n(\mathbb{P}_n)$ as $R$ goes to infinity.
The underlying idea of the bootstrap is to estimate the unknown sampling distribution $Q_n$ of the root function by this empirical distribution $\hat{Q}_{n,R}$, and estimate the measure $\xi(Q_n)$ by the plug-in estimator $\xi(\hat{Q}_{n,R})$.

In conventional bootstrap, each resample contains an average of $0.63n$ distinct sample points  --- the computational
 cost of calculating each resample estimate $\hat{\theta}(\mathcal{X}^{*}_n)$ is therefore comparable to those of the original sample.
Running $R$ iterations requires performing this task $R$ times, which can be computationally demanding for massive datasets, particularly when it involves computation of complex statistics.
This limits the application of bootstrap for massive datasets.

For BLB, we fix a subset size $b$ (typically $b = n^{\gamma}$ for some $0<\gamma < 1$) and construct a suitable number ($S$) of random subsets, $\mathcal{X}^{*j}_{n,b} = \{X_{j_1},\ldots,X_{j_b} \}, j=1,\ldots,S$, from the observed sample $\mathcal{X}_n$.
Next, for each subset $\mathcal{X}^{*j}_{n,b}$, we generate $R$ weighted resamples of size $n$ --- this can be represented by $(\mathcal{X}^{*j}_{n,b}, \mathcal{W}^{*(j,k)}_{n,b}), k=1,\ldots,R$ where $\mathcal{W}^{*(j,k)}_{n,b} = \{W_1,\cdots, W_b\}$ is a vector representing the frequencies of $(\mathcal{X}^{*j}_{n,b})$ in the $k^{th}$ resample.
The weight vector $\mathcal{W}^{*(j,k)}_{n,b}$ is generated from a multinomial distribution with $n$ trials and $b$ cells with uniform chance for each cell, independently of the subset.
Treating $\hat{\theta}(\mathcal{X}^{*j}_{n,b})$ as the population parameter and the resample estimate $\hat{\theta}(\mathcal{X}^{*j}_{n,b}, \mathcal{W}^{*(j,k)}_{n,b})$ as an estimated value of this parameter, we compute the root $T_n(\hat{\theta}(\mathcal{X}^{*j}_{n,b}, \mathcal{W}^{*(j,k)}_{n,b}), \hat{\theta}(\mathcal{X}^{*j}_{n,b}))$ for each resample, and obtain the empirical distribution $\hat{Q}^j_{n,b,R}$ of this ensemble of roots.
In the same spirit as bootstrap, we apply the plug-in estimator $\xi(\hat{Q}^j_{n,b,R})$ for each subset, and average over different subsets to obtain the estimator $\frac{1}{S} \sum_{j=1}^{S}\xi(\hat{Q}^j_{n,b,R})$ of $\xi(Q_n)$.

We propose a subsampled double bootstrap scheme (SDB) based on subsets in the same manner as BLB, but using only one resample per subset.
We fix a subset size $b$ and construct a large number ($S$) of random subsets, $\mathcal{X}^{*j}_{n,b} = \{X_{j_1},\ldots,X_{j_b} \}, j=1,\ldots,S$, from the observed sample $\mathcal{X}_n$.
However, we generate only one resample from the $j^{th}$ subset, corresponding to $\mathcal{W}^{*(j,1)}_{n,b}$ as defined above, and calculate a single root $T^{*j}_n = T_n(\hat{\theta}(\mathcal{X}^{*j}_{n,b}, \mathcal{W}^{*(j,1)}_{n,b}), \hat{\theta}(\mathcal{X}^{*j}_{n,b}))$ from the resample estimate and the subset estimate.
With this ensemble  $\{R^{*j}_n: j=1,\ldots,S\}$, we compute the empirical distribution $\hat{Q}_{n,b,S}$ of roots and
estimate $\xi(Q_n)$ using the plug-in estimator $\xi(\hat{Q}_{n,b,S})$.
Algorithm \ref{algo-new} outlines the computational steps involved.

Note that the computational advantages of SDB and BLB relative to the bootstrap are applicable only when the estimator $\hat{\theta}$ of interest can take the weighted data representation as its argument.
This property holds for a large class of commonly used estimators, including $M$-estimators.
For a subset $\mathcal{X}^{*}_{n,b}$ with resample weights $\mathcal{W}^{*}_{n,b}$, the resample estimate can then be expressed as $\hat{\theta}(\mathcal{X}^{*}_{n,b}, \mathcal{W}^{*}_{n,b})$.
Since BLB and SDB resamples have nominal size $n$ but only $O(b)$ distinct points, computing the resample estimate for these methods is much cheaper than that for bootstrap, which has $O(n)$ distinct points in the resamples.

\begin{algorithm}[h]
\SetKwInOut{Input}{Input}
\SetKwInOut{Output}{Output}
\caption{SDB algorithm}
\begin{multicols}{2}
\Input{Data $\mathcal{X}_n = \{X_1,\ldots,X_n \}$\\
$\theta$: parameter of interest \\
$ \hat{\theta}_n $: estimator\\
$T_n(\hat{\theta}_n, \theta)$: root function \\
$\xi(\cdot)$: measure of accuracy\\
$b$: subset size\\
$S$: number of subsets\\}
\end{multicols}
\Output{$\xi(\hat{Q}_{n,b,S})$: Estimate of $\xi$}
\BlankLine
\For{$j\leftarrow 1$ \KwTo $S$}{
(i) Choose random subset $\mathcal{X}^{*j}_{n,b}$ from $\mathcal{X}_n$\\
(ii) Compute $\hat{\theta}(\mathcal{X}^{*j}_{n,b})$ from $\mathcal{X}^{*j}_{n,b} $\\
(iii) Generate resample $(\mathcal{X}^{*j}_{n,b}, \mathcal{W}^{*(j,1)}_{n,b})$ from $\mathcal{X}^{*j}_{n,b}$\\
(iv) Compute resample estimate $\hat{\theta}(\mathcal{X}^{*j}_{n,b}, \mathcal{W}^{*(j,1)}_{n,b})$ from $(\mathcal{X}^{*j}_{n,b}, \mathcal{W}^{*(j,1)}_{n,b}) $\\
(v) Compute resample root: $T^{*j}_n = T_n(\hat{\theta}(\mathcal{X}^{*j}_{n,b}, \mathcal{W}^{*(j,1)}_{n,b}), \hat{\theta}(\mathcal{X}^{*j}_{n,b}))$\\
}
1. Compute empirical distribution of roots: $\hat{Q}_{n,b,S} = \text{ ecdf of } \{T^{*1}_n, \ldots, T^{*S}_n\}$\\
2. Calculate the plug-in estimator $\xi(\hat{Q}_{n,b,S})$
\label{algo-new}
\end{algorithm}

\subsection{Comparison of SDB, BLB, and Bootstrap}
\label{sec:comparison}
For both BLB and SDB, the resample estimation step applies to a resample with $O(b)$ distinct points, whereas in bootstrap the resample has $O(n)$ distinct points.
This makes SDB and BLB computationally much cheaper than bootstrap when $b << n$.
Denote the computational time for performing the estimation process $\hat{\theta}$ on a sample of size $m$ by $t(m)$.
In this formulation of computational time we focus on sample size to illustrate the resampling methods, and ignore other factors affecting computational time.
For an estimator that can take the weighted data representation, the estimation time for a resample with nominal size $n$ but only $b$ distinct points is $t(b)$.
 The estimation time for bootstrap, BLB, and SDB, for conducting inference in one original data sample, are listed in Table \ref{tab-time}, where the symbols have the same meaning as earlier.
Bootstrap requires estimation on the original data and its $R$ resamples.
Each BLB subset requires estimation on the subset and its $R$ resamples.
Each SDB subset requires estimation on the subset and the single resample.

\begin{table} [h!]
\centering
\begin{tabular}{|c|c|}
\hline Name & Estimation Time\\
\hline Bootstrap & $(R+1)\times t(n)$ \\
\hline BLB & $ S(R+1)\times t(b)$ \\
\hline SDB & $2S \times t(b)$ \\
\hline
\end{tabular}
\caption{Estimation time for different resampling methods}
\label{tab-time}
\end{table}

For BLB, \cite{kleiner2014scalable} recommends R = 100 and a small value of S (2-10 depending on $b$).
For illustration, let $n=100,000$ and $b=n^{0.6}$, then the number of distinct points in each resample is at most $1000$, resulting in much faster computation than bootstrap.
However, in terms of sample coverage, each subset can cover at most $1\%$ of the data, so 10 subsets can at best cover 10\% of the data at an expense of  1010$\times t(b)$.
The SDB can run more than 500 different subsets at the same expense, providing a far more comprehensive coverage of the data.

Further, given a certain time budget, it is not clear how to choose the tuning parameters $R$ and $S$ that will provide optimal statistical accuracy for BLB.
The adaptive method proposed by \cite{kleiner2014scalable} provides an interesting alternative by choosing a tolerance parameter $\epsilon$ instead of $R$ and $S$.
But even then, it is not clear how to choose an appropriate $\epsilon$ in practice, since $t(m)$ is not known a priori, and neither do we know the estimation variability as a function of sample size.
For the SDB (with a given subset size) and the bootstrap, the estimation time involves only one parameter, the number of resamples (or subsets), and hence the practitioner can simply keep running resamples until the time budget runs out.

\section{Theory for independent data}
\label{sec:ind_theory}

In this section, we provide a theoretical analysis  of the SDB in a general empirical process setting. Consider a class of functions $\Fc$ [each element mapping from $\R^k$ to $\R$]. Denote by $\ell^\infty(\Fc)$ the space of bounded functions which map from $\Fc$ to $\R$. To describe consistency of the SDB, consider the \textit{SDB-process}
\[
\hat \GG_{n,b}^B(f) := \frac{1}{\sqrt{n}}\sum_{i=1}^b (W_{i,n}-n/b)f(X_{R^{-1}(i)}).
\]
Here, $W := (W_{1,n},...,W_{b,n}) \sim \text{Multinomial}_b(n,1/b,...,1/b)$ independent of $X_1,...,X_n$ and $R$ follows a uniform distribution on the permutations of $\{1,...,n\}$ and is independent of $X_1,...,X_n,W$. Note that in empirical process settings, it is important to specify the underlying probability space. This is done in the mathematical appendix [see Section \ref{sec:proof1}]. In order to show that the SDB `works' in a process setting, we need to establish that the distribution of the SDB process $\hat \GG_{n,b}^B$ [conditional on the observations $X_i$] is close to the distribution of the empirical process $\GG_{n}$ where
\[
\GG_{n}(f) := \frac{1}{\sqrt{n}}\sum_{i=1}^n (f(X_i) - \E[f(X_1)])
\]
when both are viewed as elements of $\ell^\infty(\Fc)$. To this end we show that the SDB-process converges in distribution, conditionally on the data $X_1,...,X_n$, to the same Gaussian process as the empirical process $\GG_n$.

\begin{theorem} \label{th:SDBiid}
Assume that $\Fc$ is a Donsker class for $P$, that $X_i\sim P$ are i.i.d. and that additionally $\Fc_\delta := \{f-g: f,g\in \Fc, P(f-g)^2\ \leq \delta\}$ is measurable in the sense discussed in \cite{ginzin1984} for each $\delta>0$. Then we have for $\min(n,b)\to\infty$
\[
\hat \GG_{n,b}^B \weakP{W,R} \GG
\]
in $\ell^\infty(\Fc)$ where $\GG$ denotes a centered Gaussian process with covariance $\E[ \GG(f)\GG(g)] = \cov(f(X),g(X))$.
\end{theorem}

In the above Theorem, conditional weak convergence $\weakP{W,R}$ is in the sense described in \cite{kos2008}, Section 2.2.3. A proof of this result can be found in the mathematical appendix [Section \ref{sec:proof1}].

One remarkable fact about Theorem \ref{th:SDBiid} is that, in addition to $\Fc$ being P-Donsker, the only requirement on the class of functions $\Fc$ is a mild measurability condition. This means that the SDB on a process level `works' whenever the corresponding functional central limit Theorem holds true (up to the mild measurability assumption on the function class $\Fc$), i.e. the SDB can be applied in a very wide variety of settings. The proof relies on basic tools from empirical process theory [in particular, a fundamental result on the exchangeable bootstrap, see Theorem 3.6.3 in \cite{vanwel1996}], but is completely different from the proof of Theorem 1 in \cite{kleiner2014scalable}. The reason is that in the BLB one initial subset is fixed, while in the SDB a different subset of the data is used in each iteration. The latter fact poses additional challenges for the theoretical analysis of SDB.

Theorem \ref{th:SDBiid} provides a fundamental building block for the analysis of SDB.  Combined with the continuous mapping theorem and functional delta method for the bootstrap [see for instance \cite{kos2008}, Theorem 10.8 and Theorem 12.1], it can be utilized to validate the consistency of SDB for a wide range of applications. For illustration purposes, let us consider an application of the functional delta method for the bootstrap with the root $T_n(\hat \theta_n,\theta) := \sqrt{n}(\hat \theta_n - \theta)$. Assume that we are interested in conducting inference on a parameter $\theta$ which can be represented as $\phi((f \mapsto P f)_{f \in \Fc})$, and the estimator takes the form $\hat\theta(\Xc) = \phi((f \mapsto \p_n f)_{f \in \Fc})$
for a suitable map $\phi$. More precisely, we assume that $\phi$ satisfies the following condition
\begin{enumerate}
\item[(H)] There exists a $V$ which is a vector space with $V \subset \ell^\infty(\Fc)$ such that the sample paths of $\GG$ lie in $V$ with probability one. The map $\phi: \ell^\infty(\Fc) \to \R^k$ is compactly differentiable tangentially to $V$ in the point $H: f \mapsto P f$. Denote the corresponding derivative by $\phi_H'$.
\end{enumerate}
For $f \in \Fc$, write $\p_{n,b} f := \frac{1}{n}\sum_{i=1}^b W_{i,n} f(X_{R^{-1}(i)})$. Then, in the notation from Section \ref{sec:ind}, we have $\hat\theta(\Xc_{n,b}^{*j}, \mathcal{W}_{n,b}^{*(j,1)}) = \phi((f \mapsto \p_{n,b} f)_{f \in \Fc})$. Now the delta method for the bootstrap [Theorem 12.1 in \cite{kos2008}] yields for $\min(n,b) \to \infty$
\[
T_n(\hat\theta(\Xc_{n,b}^{*j},\mathcal{W}_{n,b}^{*(j,1)}),\hat\theta(\Xc_n)) = \sqrt{n}(\hat\theta((\Xc_{n,b}^{*j},\mathcal{W}_{n,b}^{*(j,1)})) - \hat\theta(\Xc_n)) \weakP{W,R} \phi_H' \GG.
\]
At the same time, the classical functional delta method yields
\[
T_n(\hat\theta(\Xc_n),\theta) = \sqrt{n}(\hat\theta(\Xc_n) - \theta) \weak \phi_H' \GG.
\]
Assume measurability of $\hat\theta((\Xc_{n,b}^{*j},\mathcal{W}_{n,b}^{*(j,1)})), \hat\theta(\Xc_n)$. Write $\mathcal{L}$ for the distribution of $\phi_H' \GG$, $\mathcal{L}_n$ for the distribution of $T_n(\hat\theta(\Xc_n),\theta)$, and denote by $\mathcal{L}_{n,b}^B(R,W)$ the distribution of $T_n(\hat\theta(\Xc_{n,b}^{*j},\mathcal{W}_{n,b}^{*(j,1)}), \hat\theta(\Xc_n))$ conditional on $R,W$. Denoting by $\mathrm{d}$ a metric on the space of distributions on $\R^k$ which metrizes weak convergence, we have proved that $\mathrm{d}(\mathcal{L}_n,\mathcal{L}) \to 0$ as $n \to \infty$ and $\mathrm{d}(\mathcal{L}_{n,b}^B(R,W),\mathcal{L}) \to 0$ in outer probability as $\min(n,b) \to \infty$. In particular, this shows that for any map $\xi$ from the space of distributions to $\R^k$ which is continuous in the point $\mathcal{L}$ with respect to the metric $\mathrm{d}$, we have $\xi(\mathcal{L}_{n,b}^B(R,W)) - \xi(\mathcal{L}) \to 0$ in outer probability. This shows that the conclusion of Theorem 1 in \cite{kleiner2014scalable} continues to hold in the SDB setting.


\section{Simulation study for independent data}
\label{sec:ind_sim}
{In this section, we report two simulation studies comparing the performance of bootstrap, BLB, and SDB in large simulated datasets in the i.i.d. framework.}
We used model settings similar to \cite{kleiner2014scalable}.
Since they have already demonstrated that BLB performs better than the $m$ out of $n$ bootstrap and subsampling, we did not include these methods in our study.

\subsection{Multiple Linear Regression}
\label{sec:ind_sim_reg}
Consider a $d$-dimensional multiple linear regression model
\begin{equation*}
y_i = \beta_1 x_{i,1}+ \ldots+\beta_d x_{i,d}+e_i
\end{equation*}
for $i = 1, \ldots,n$.
Our parameter of interest is the $d$-dimensional vector of slope coefficients, {whose true value is $\beta = (\beta_1,...,\beta_d) = (1,\ldots,1)'$}.
We use the usual OLS estimator $\hat{\beta}$.
We also want to construct a simultaneous 95\% confidence region for $\beta$.
Traditionally we use the F-statistic
\begin{equation*}
T_n(\hat{\beta},\beta) =
\frac{(\hat{\beta}-\beta)'X'X(\hat{\beta}-\beta)/d}
{(y-X\hat{\beta})'(y-X\hat{\beta})/(n-d-1)}
\end{equation*}
to construct the joint confidence region.
Let $q_{0.95}$ be the 95\% quantile of the true (unknown) distribution of $T_n(\hat{\beta},\beta)$.
Then the confidence region is given by $\{\beta: T_n(\hat{\beta},\beta) \le q_{0.95} \} $.
In general the true distribution of $T_n$, and hence its quantile $q_{0.95}$, is unknown.
But it can be estimated by the resampling techniques described in the previous section, with $\xi(Q_n) = q_{0.95}$ where $Q_n$ is the true distribution of $T_n$.

In our simulations, we use a model from the simulation study of \cite{kleiner2014scalable}.
We generate $x_{i,j} \stackrel{iid}{\sim} t_3$ and $e_i \stackrel{iid}{\sim} N(0,100)$ independently.
For normally distributed errors, we know that $T_n \sim F(d, n-d-1)$,
and hence the true quantiles are given by those of the corresponding $F$ distribution.
We define the error rate as
\begin{equation*}
|\frac{\hat{q}_{0.95}}{q_{0.95}}-1|
\end{equation*}
where
$\hat{q}$ and $q$ represent the estimated and true quantiles of $T_n$, respectively.
{We use subset size $b = n^\gamma$ with $\gamma = 0.6, 0.7, 0.8$ 
for both BLB and SDB, and let
$n$=100000, $d$=100.}
Following \cite{kleiner2014scalable} we fix $R=100$ for BLB.
We allowed the competing methods to run for 60 seconds.

\subsection{Logistic Regression}
\label{{sec:ind_sim_logreg}}
Consider a $d$-dimensional multiple logistic regression model
\begin{equation*}
y_i \stackrel{ind}{\sim} Ber(p_i) \text{ where } p_i = \beta_1 x_{i,1}+ \ldots+\beta_dx_{i,d}
\end{equation*}
for $i = 1, \ldots,n$.
Our parameter of interest is the $d$-dimensional vector of slope coefficients, {whose true value is $\beta = (\beta_1,\ldots,\beta_d) = (1,\ldots,1)'$}.
We use the maximum likelihood estimator $\hat{\beta}_n$ which does not have a closed form expression for this model, but can be numerically computed using a Newton-Raphson method.
We use the R function \emph{glm} for fitting the model.
As before, we want to construct a simultaneous 95\% confidence region for $\beta$.
Define the root function
\begin{equation*}
T_n(\hat{\beta},\beta) =
(\hat{\beta} - \beta)' \hat{\Sigma} (\hat{\beta} - \beta)
\end{equation*}
where $\hat{\Sigma} = \sum_{i=1}^{n} \frac{\exp(x_i'\hat{\beta})}{[1+\exp(x_i'\hat{\beta})]^2} x_i x_i'$.
Let $q_{0.95}$ be the 95\% quantile of the true (unknown) distribution of $T_n(\hat{\beta},\beta)$.
Then the confidence region is given by
$ \{\beta: T_n(\hat{\beta},\beta) \le q_{0.95} \} $.
In general the true distribution of $T_n$, and hence its quantile $q_{0.95}$, is unknown.
But it can be estimated by the resampling techniques described in the previous section, with $\xi(Q_n) = q_{0.95}$ as the target precision parameter, where $Q_n$ is the true distribution of $T_n$.

We generate $x_{i,j} \stackrel{iid}{\sim} t_3$ and
obtain a numerical approximation of $q_{0.95}$ using 10000 Monte Carlo simulations.
As before, we define the error rate as
$|{\hat{q}_{0.95}}/{q_{0.95}}-1| .$
{We use subset size $b = n^\gamma$ with $\gamma = 0.6, 0.7, 0.8$ 
for both BLB and SDB, and use $n$=100000, $d$=10.}
 Following \cite{kleiner2014scalable} we fix $R=100$ for BLB.
We allowed the competing methods to run for 20 seconds.

\subsection{Comparison of SDB, BLB, and Bootstrap}
\label{sec:iid_sim_comparison}
The methods are compared with respect to the time evolution of error rates.
Note that this is different from conventional analysis where error rates from competing methods are compared for the same number of iterations.
This makes sense because different methods have different estimation time profiles (as formulated in Table \ref{tab-time}), and we want to investigate which method is the fastest to produce reasonably accurate results.
We consider a time grid $1, 2, \ldots, 60$ (in seconds) and at each time point $t$, we look up the latest iteration that was completed by this time, and calculate the error corresponding to the estimate $\hat{\xi}$ from cumulative iterations including that iteration.
For each method and any $t$, this can be interpreted as the error rate obtained by that method for a given computation time budget of $t$ seconds.
Different methods will have different numbers of iterations completed within the same time budget.
For bootstrap and SDB, latest iteration means the latest completed resample or subset-resample, while for BLB (following \cite{kleiner2014scalable}) the latest iteration means the latest completed subset.
Note that till the first iteration is complete, we do not have an estimate $\hat{\xi}$, so we consider the error rate to be 1 till the first iteration is completed.
Error rates are averaged across 20  Monte Carlo simulations.

\begin{figure}[!htb]
    \centering
    \includegraphics[height=0.25\textheight]{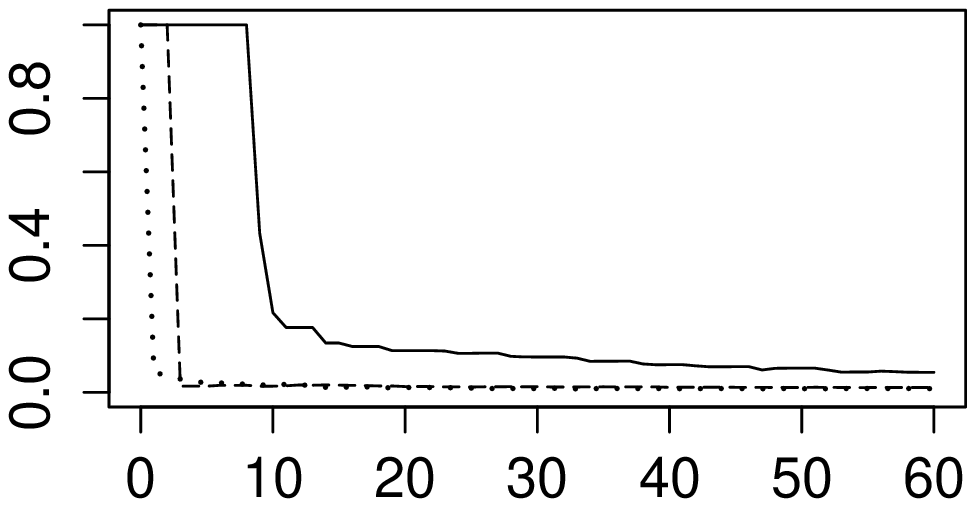}
    \includegraphics[height=0.25\textheight]{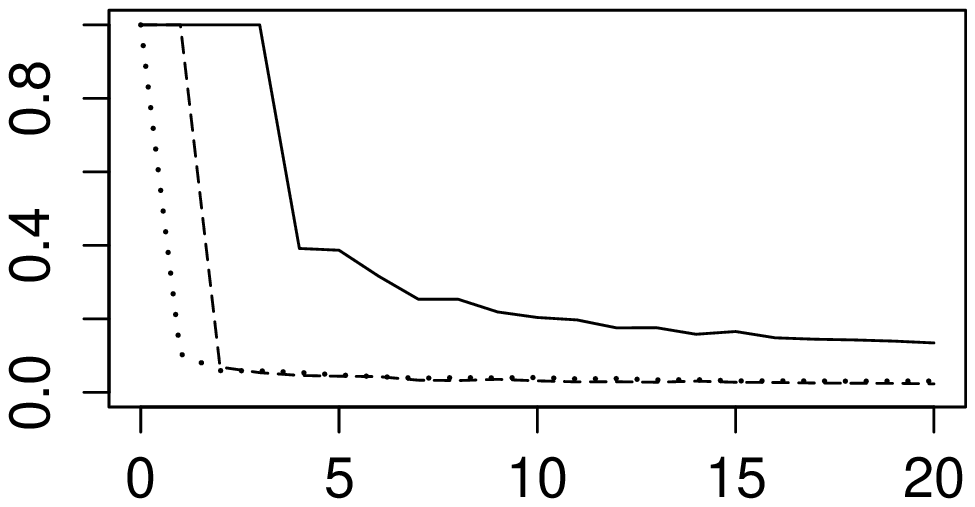}
    \includegraphics[height=0.25\textheight]{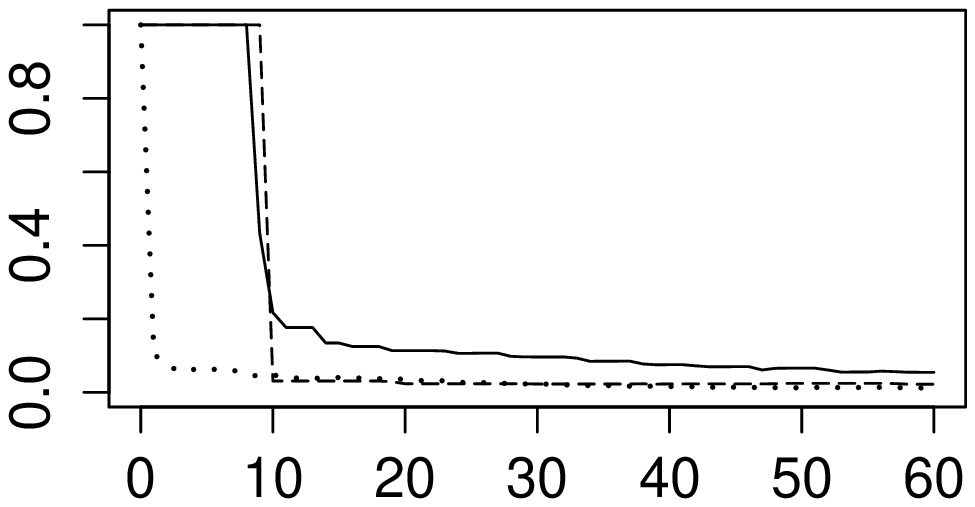}
    \includegraphics[height=0.25\textheight]{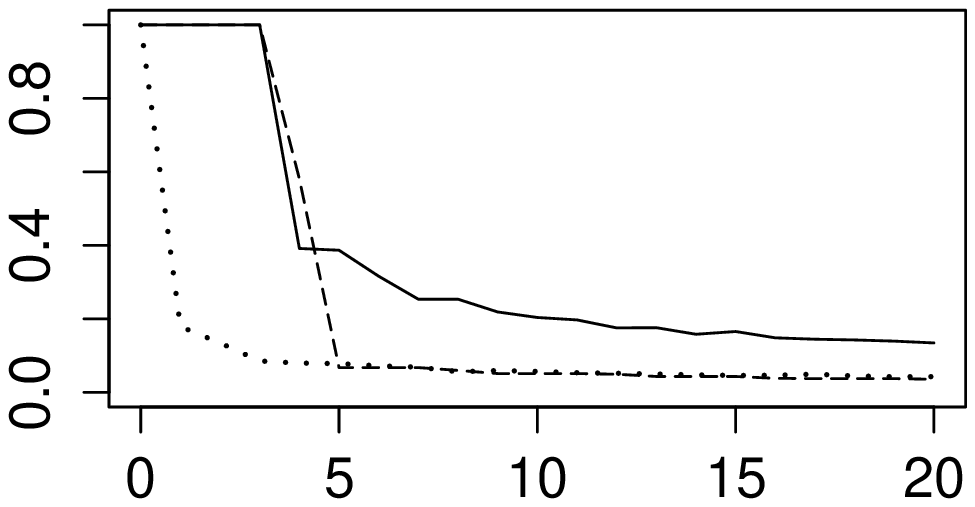}
    \includegraphics[height=0.25\textheight]{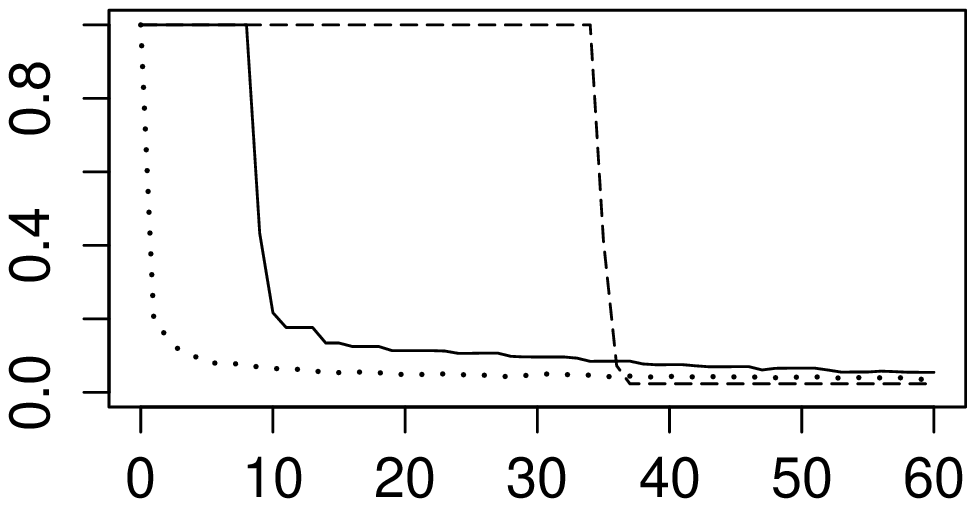}
    \includegraphics[height=0.25\textheight]{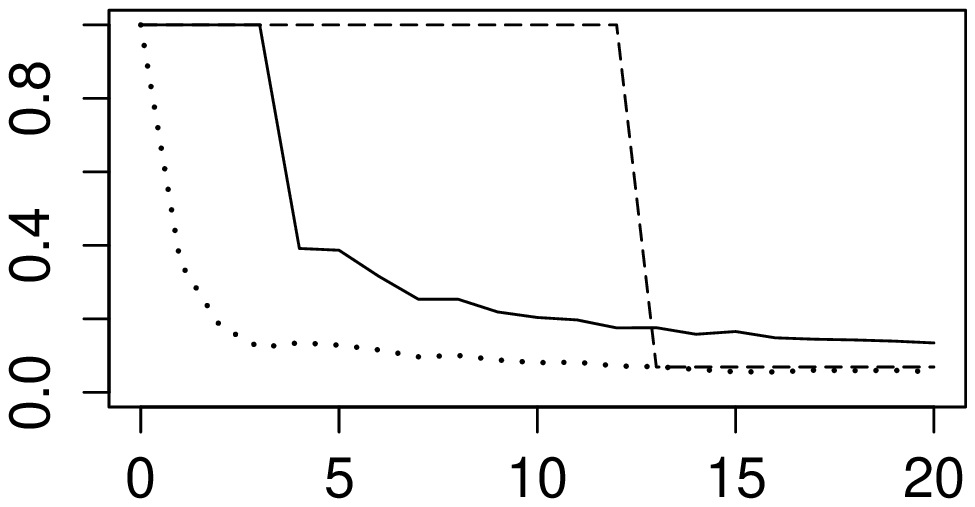}
    \caption{Time evolution of error rates for multiple linear regression with $d$=100 (left column) and multiple logistic regression with $d$=10 (right column).
    Sample size $n$=100000, subset size  is $b=n^\gamma$ where $\gamma = 0.6$ (top row), $\gamma = 0.7$ (middle row) and $\gamma = 0.8$ (bottom row).
  Bootstrap errors are represented by solid lines, BLB errors by dashed lines, and SDB errors by dotted lines.
 Errors are averaged over 20 simulations.}
 \label{simnew}
\end{figure}

Figure \ref{simnew} shows the time evolution of error rates for bootstrap, BLB, and SDB.
Bootstrap has the highest computing cost which gets reflected in its slow convergence.
The performance of BLB and SDB are close to one another for generous time budgets, but for lower time budgets SDB performs better by quickly giving a reliable estimate while BLB takes some time to complete the first subset.
This phenomenon becomes particularly prominent for higher values of $b$ as BLB's computing time for each subset becomes large.
For small time budgets even bootstrap can beat BLB when $b = n^{0.8}$, since the time taken by BLB to complete a subset can exceed the given budget.
A similar phenomenon for small time budgets was observed in the simulation study of \cite{kleiner2014scalable} (see Figure 1(a)---(c)  in their paper for linear regression and 2 (a)---(c) for logistic regression), where bootstrap estimates are available but BLB estimates are not available yet for subset size $b = n^{0.8}$ or $b = n^{0.9}$.

\begin{remark}
\normalfont
Computing time for the resampling methods depends on various aspects of the computational infrastructure used, e.g. the processing power of the computer, storage capacity, and statistical software or computing platform.
All our simulations were performed on a desktop computer with Intel(R) Core(TM)2 Duo CPU E8400 @3.00 GHz processor and 4 GB RAM, running R version 3.0.1.
The computational infrastructure influences the computing time of various resampling methods in identical manner, so the relative performance of these methods should be qualitatively similar in a different infrastructure, even if the absolute performances might vary.
\end{remark}

\begin{remark}
\normalfont
It is relevant to note that while we have used models from \cite{kleiner2014scalable} in these studies, the precision measure and the method of comparison between resampling schemes are slightly different.
They constructed marginal confidence intervals for the individual regression coefficients, and combined results for the $d$ coefficients by averaging the error rate over the dimensions.
Thus they are interested in measures of precision of the individual estimation tasks of estimating the $d$ coefficients.
However, in a multivariate regression setting, the joint estimation task of all coefficients taken together might be of more interest.
Accordingly, we constructed a simultaneous confidence region for the $d$-dimensional vector of regression coefficients to assess precision of this joint estimation task, and compute error in terms of this confidence region.

For comparison between resampling schemes, \cite{kleiner2014scalable} allowed the competing resampling schemes to converge, and
for each iteration (defined as a complete subset for BLB and resample for bootstrap), they computed the average cumulative computing times and average error rates from five Monte Carlo simulations.
They compared resampling schemes on the basis of this average time vs average error trade-off.
In our simulations, we compare methods on the basis of error rate achieved for a given time budget, over 20 Monte Carlo simulations.
Thus time is not averaged across simulations --- rather, at a fixed point in time we look up the error rates obtained by this time in the Monte Carlo simulations, and average them.
If at a certain time point no estimate is available yet (no iteration has been completed), we assign an error rate of 1.

In particular, in our simulation plots, the error rate changes only when an iteration has been completed.
This makes them look `jerky' and unstable as there are long stretches of a flat line followed by a sudden drop.
Since estimates change only upon the completion of a new iteration, the arrival of new estimates is actually an intermittent process rather than a continuous process with respect to the time axis, and the error rate does not change unless a new estimate is available.
Therefore it is realistic that error rates change in a `jerky' fashion rather than smoothly, and this is not a symptom of instability.
\end{remark}

\begin{remark}
\normalfont
Several bootstrap approaches exist in a regression setting ---  for example paired bootstrap, residual bootstrap, wild bootstrap and so on.
In these simulations we have implemented the paired bootstrap, where both regressors and response are resampled.
The paired bootstrap method can be naturally extended to the BLB and SDB algorithms, but it is unclear whether there are straightforward extensions for residual bootstrap or wild bootstrap.

As pointed out by a referee, another alternative is to look at bootstrap p-values instead of confidence regions for regression models.
In our formulation $\xi$ is a parameter associated with the sampling distribution $Q_n$ of the root function $T_n$ (which in this case is the F-statistic), while the p-value is a statistic.
However, one can implement Algorithm \ref{algo-new} to `estimate' the true p-value using the empirical distribution  $\hat{Q}_{n,b,S}$.
Limited (unreported) simulation results suggest that SDB still possesses the same advantage over BLB and bootstrap, which is reported for confidence region.
A more careful investigation regarding the suitability of SDB for approximation of the p-value in theory and finite sample simulations is left for future work.


\end{remark}

\afterpage{\clearpage}

\section{SDB for time series data}
\label{sec:ts}
In this section, we extend SDB to time series data.
Note that \cite{kleiner2014scalable} have briefly mentioned an extension of BLB to the time series setting using stationary bootstrap (\cite{politis1994stationary}), however no rigorous theory is provided.
Also see \cite{laptevboot} for a recent implementation on a large Twitter dataset.

{Suppose we observe $\mathcal{X}_n = \{X_t\}_{t=1}^n$, which is a stretch of length $n$ from the strictly stationary time series $\{X_t\}_{t\in \mathbb{Z}}$, and let $P$ denote the joint probability law that governs the stationary sequence. Let $\theta = \theta(P)$ be our parameter of interest, and suppose we have an estimator $\hat{\theta}_n(\mathcal{X}_n)$ which is a measurable mapping from $\mathcal{X}_n$ to $\mathbb{R}$.}
As with independent data, we are interested in evaluating the precision of this statistical inference.
As before, this can be formulated in terms of a root function $T_n(\hat{\theta}_n, \theta)$, and the precision can be expressed as $\xi(Q_n)$ where $Q_n$ is the true (unknown) distribution of $T_n$.

For BLB, we first construct subsets of the original sample by randomly choosing a continuous stretch of data $\mathcal{X}^*_{n,b} = \{X_{J+i}\}_{i=1}^b$ where $0 \le J \le n-b+1 $ and the subset size $b$ is fixed beforehand.
From the $j^{th}$ subset, we then construct R weighted resamples $(\mathcal{X}^{*j}_{n,b}, \mathcal{W}^{*(j,k)}_{n,b})$ for $k=1,\ldots,R$ of size $n$ using the Moving Block Bootstrap (MBB, hereafter) of \cite{kunsch1989jackknife} and \cite{liu1992moving}.
We use MBB instead of stationary bootstrap used by \cite{kleiner2014scalable} since the MBB is conceptually simpler, and easier in terms of theoretical treatment.
For this, we consider some suitable block length $L < b$ and divide the subset $\{ X_J, X_{J+1}, \ldots, X_{J+b-1}\}$ into an ensemble of overlapping blocks $\{ X_i, X_{i+1}, \ldots, X_{i+L-1}\}$ where $J \le i \le J+b-L+1$.
The resample is constructed by concatenating blocks that are randomly sampled from this ensemble, till we obtain a chain of size $n$.
Note that when $n$ is not a multiple of $L$, we will need to take a fraction of the final block in order to obtain a resample of length exactly $n$.
This gives us an ensemble of roots of the form $T_n(\hat{\theta}(\mathcal{X}^{*j}_{n,b}, \mathcal{W}^{*(j,k)}_{n,b}),  \hat{\theta}_b(\mathcal{X}^{*j}_{n,b})), k = 1, \ldots, R$, and we use the empirical distribution of this ensemble to approximate the unknown distribution of $T_n(\hat{\theta}_n, \theta)$. 
Averaging over $j=1,\ldots,S$ subset estimates then gives the BLB estimate of precision.

For SDB, for each subset $\mathcal{X}^{*j}_{n,b}$, we generate only one MBB resample $(\mathcal{X}^{*j}_{n,b}, \mathcal{W}^{*(j,1)}_{n,b})$ to construct the root $T_n(\mathcal{X}^{*j}_{n,b}, \mathcal{W}^{*(j,1)}_{n,b}),  \hat{\theta}_b(\mathcal{X}^{*j}_{n,b}))$.
We do this a large number ($S$) of times to generate an ensemble of roots, and use the empirical distribution of this ensemble to approximate the unknown distribution of $T_n(\hat{\theta}_n, \theta)$.
Algorithm \ref{algo-ts} outlines the computational steps involved.

\begin{algorithm}[h!]
\SetKwInOut{Input}{Input}
\SetKwInOut{Output}{Output}
\caption{SDB time series algorithm}
\begin{multicols}{2}
\Input{Data $\mathcal{X}_n = \{X_1,\ldots,X_n \}$\\
$\theta$: parameter of interest \\
$ \hat{\theta}_n $: estimator\\
$T_n(\hat{\theta}_n, \theta)$: root function \\
$\xi(\cdot)$: measure of accuracy\\
$b$: subset size\\
$L$: block length\\
$S$: number of subsets\\}
\end{multicols}
\Output{$\xi(\hat{Q}_{n,b,S})$: Estimate of $\xi$}
\BlankLine
\For{$j\leftarrow 1$ \KwTo $S$}{
(i) Choose random subset $\mathcal{X}^{*j}_{n,b} =  \{X_{J+i}\}_{i=1}^b$ from $\mathcal{X}_n$ where $0 \le J \le n-b+1 $\\
(ii) Compute $\hat{\theta}(\mathcal{X}^{*j}_{n,b})$ from $\mathcal{X}^{*j}_{n,b} $\\
(iii) Choose $k = n/L$ blocks by randomly sampling $k$ starting points $(t_1, \ldots,t_k)$ from $\{J+1,\ldots,J+b-L+1\} $ with replacement\\
(iv) Construct resample weights for subset:\\
 Initialize: $\mathcal{W}^{*(j,1)}_{n,b} \leftarrow (\underbrace{0 \ldots 0}_{b})$\\
\For{$i\leftarrow 1$ \KwTo $k$} {
					$\mathcal{W}^{*(j,1)}_{n,b} \leftarrow \mathcal{W}^{*(j,1)}_{n,b} + (
					\underbrace{0 \ldots 0}_{t_i - 1} \; \underbrace{1 \ldots 1}_{L}  \;
					\underbrace{0 \ldots 0}_{b-t_i-L+1}
					)$
									}
(v) Compute resample estimate $\hat{\theta}(\mathcal{X}^{*j}_{n,b}, \mathcal{W}^{*(j,1)}_{n,b})$ from $(\mathcal{X}^{*j}_{n,b}, \mathcal{W}^{*(j,1)}_{n,b}) $\\
(vi) Compute resample root: $R^{*j}_n = T_n(\hat{\theta}(\mathcal{X}^{*j}_{n,b}, \mathcal{W}^{*(j,1)}_{n,b}), \hat{\theta}(\mathcal{X}^{*j}_{n,b}))$\\
}
1. Compute empirical distribution of roots: $\hat{Q}_{n,b,S} = \text{ ecdf of } \{R^{*1}_n, \ldots, R^{*S}_n\}$\\
2. Calculate the plug-in estimator $\xi(\hat{Q}_{n,b,S})$
\label{algo-ts}
\end{algorithm}

In the time series case,
estimation time can be formulated as $t(m)$ in a manner similar to Section \ref{sec:comparison}, where $m$ is the number of distinct points, and the estimation times listed in Table \ref{tab-time} apply with MBB taking the place of bootstrap.
MBB requires estimation on the original data and its $R$ resamples.
Each BLB subset requires estimation on the subset and its $R$ resamples.
Each SDB subset requires estimation on the subset and the single resample.

%

Broadly speaking, the time series version of SDB retains the advantages discussed in Section \ref{sec:comparison} in the context of independent data.
For a given computational time budget, by using a single resample for each random subset SDB can accommodate much more comprehensive coverage of data than BLB.
Also, BLB involves tuning parameters $R$ and $S$ (or $\epsilon$ under the adaptive method) whose selection can be non-trivial, while SDB and MBB do not require this type of tuning parameter selection.
However, an important tuning parameter in the time series setting is the block length $L$ which can affect both variability and the accuracy of the estimate of precision in all three resampling methods.

\section{Theory for dependent data}
\label{sec:ts_theory}
We begin by setting up a mathematical framework for SDB in the dependent case. Throughout this section we assume that the observations $X_1,...,X_n$ stem from a strictly stationary time series $\{X_t\}_{t \in \Z}$. Given a sample $X_1,...,X_n$, the SDB procedure for time series can be described through the following steps.
\begin{enumerate}
\item Pick a random variable $J$ which is distributed uniformly on $0,....,n-b-1$. This corresponds to the first step of randomly selecting a block of length $b$ from the complete data.
\item Choose $K = \lceil n/L \rceil$ random variables $s_1,...,s_K$ which are i.i.d. and distributed uniformly on $0,...,b-L+1$. Generate the sample $X_1^*,...,X_n^*$ by setting
\[
(X_{kL +1}^*,...,X_{(k+1)L}^{*}) := (X_{J+s_k},...,X_{J+s_k+L-1})
\]
\item After the first two steps above, one realization of the SDB process is given by
\[
\hat \GG_{n,b}^B(f) := \frac{1}{\sqrt{n}}\sum_{i=1}^n \Big(f(X_i^*) - \frac{1}{b} \sum_{j = 1}^{b} f(X_{J + j})\Big).
\]
Repeat a large number of times, each time generating a new $J,s_1,...,s_K$.
\end{enumerate}

As in the case of independent observations, our aim is to establish validity of the SDB for general classes of functions. In contrast to the i.i.d. setting, where the `classical' bootstrap for empirical processes is well understood, there are very few results on bootstrap validity for general empirical processes. All of the available results rely on the notion of $\beta$-mixing to measure dependence. More precisely, the k'th $\beta$-mixing coefficient $\beta(k) $ with $k \in \N$ is defined as
\[
\beta(k) := \frac{1}{2} \sup \sum_{(i,j) \in I\times J} |P(A_i\cap B_j) - P(A_i)P(B_j)|
\]
where the supremum is taken over all finite measurable partitions $(A_i)_{i \in I}, (B_j)_{j \in J}$ of $\sigma(X_t,t\leq 0)$ and $\sigma(X_t,t\geq k)$, respectively. As of this writing, we are aware of only three articles that deal with bootstrap validity for general empirical processes based on dependent observations. \cite{buh1995} considers the moving blocks bootstrap under exponential decay of the $\beta$-mixing coefficients and classes of functions with polynomial bracketing numbers. \cite{rad1996} establishes the validity of the moving blocks bootstrap for VC [see \cite{vanwel1996}, Chapter 2.6.2 for a definition] classes of functions under conditions on polynomial decay of $\beta$-mixing coefficients. Finally, \cite{rad2009} revisits the disjoint blocks bootstrap and proves its validity under generic conditions on the function class and decay of $\beta$-mixing coefficients. The latter paper also contains a nice overview of literature on bootstrap validity under dependence [see also \cite{rad2002} for a review of earlier results]. Our main result can be viewed as an analogue of Theorem 1 in \cite{rad1996}.

\begin{theorem}\label{th:maindep}
Assume that $\Fc$ is a permissible [as defined on page 228-229 in \cite{kos2008}] VC class with envelope function $F$ such that $\E[F^p(X_1)] < \infty$ for some $p > 2$. Assume that the mixing coefficients $\beta$ satisfy $\beta(k) \leq k^{-q}$ for some $q > p/(p-2)$. If additionally there exist $\kappa>0, \gamma>0, 0< \rho < \frac{p-2}{2(p-1)}$ such that $b, L$ satisfy
\[
n^{-1/2} L b^\gamma = o(n^{-\kappa}), \quad L \to \infty, \quad L = O(b^\rho), \quad n^{1/2} = O(b^{(p-1)\gamma}),
\]
we have
\[
\hat \GG_{n,b}^B \weakP{J,S} \GG
\]
in $\ell^\infty(\Fc)$ where $\GG$ denotes a centered Gaussian process with covariance structure
\[
\E[ \GG(f)\GG(g)] = \sum_{t \in \Z}(\E[f(X_1)g(X_t)] - \E[f(X_1)]\E[g(X_1)]).
\]
\end{theorem}

A proof of this theorem is in the mathematical appendix [Section \ref{sec:proof2}]. Theorem \ref{th:maindep} shows that the time series version of SDB also works in a wide range of settings. In particular, the continuous mapping theorem and delta method for the bootstrap can be employed in the same fashion as discussed at the end of Section \ref{sec:ind_theory}. We conjecture that the assumptions on the dependence can be weakened if we consider more specialized classes of functions, such as indicators of rectangles which would lead to the `classical' empirical distribution function. 

\section{Simulation study for time series}
\label{sec:ts_sim}
In this section we report the numerical performance of SDB, BLB, and MBB in two simulation studies involving large time series data.

\subsection{Median of AR(1) process}
\label{sec:ts_simM}
Consider an AR(1) time series formulated as
\begin{equation*}
X_{t} = \rho X_{t-1} + e_t
\end{equation*}
of
length $n = 100,000$ and random innovation $e_t \stackrel{iid}{\sim} N(0,1) $.
The parameter of interest is the population median $M$. We define $T_n = \sqrt{n}(M_n-M)$ where $M_n$ is the sample median. {We are interested in evaluating the precision of the estimator $M_n$. }
 Our measure of precision is a quantile of the distribution of $T_n$, i.e.
$\xi = q_{\alpha}(T_n)$
which can be used for constructing confidence intervals, for example, with $\alpha=5\%, 95\% $ we can construct a $90\%$ confidence interval.

We obtain the `true' value $\xi_{true}$ from 10000 simulations.
The error rate is measured by
$ \mid {\hat{\xi}}/{\xi_{true}} - 1 \mid.$
We implement and compare the three resampling methods, namely MBB, BLB, and SDB.
Block length is $L=10,20,50$ for all methods, and we use subset sizes $b = 5000, 10000$ for BLB and SDB.
We allow each method to run for 60 seconds for $L=20,50$ and 120 seconds for $L=10$, to allow BLB to complete one subset.

\subsection{Time Series Regression}
\label{sec:ts_simreg}

We also studied the relative performance of MBB, BLB, and SDB in the time series regression framework (see e.g. \cite{andrews1992improved}, \cite{kiefer2000simple}, \cite{rho2013improving}).
Consider the time series regression model
\begin{equation*}
y_t = X'_t\beta + u_t
\end{equation*}
for $t=1,\ldots,n$ where $\beta$ is a $d\times1$ vector of regression coefficients, $X_t$ is a $d\times1$ vector of stationary regressors, and $u_t$ is a stationary error process that satisfies $\mathbb{E}[u_t \mid X_t]$ = 0.
We considered the AR(1)-HOMO regression model of \cite{andrews1992improved} where the $d$ regressors and the error process are mutually independent, mean zero, homoskedastic, AR(1) processes with autocorrelation $\rho$ and standard normally distributed innovations, and set $\beta = \mathbf{0}$.
We set $n = 100,000$ and $d=10$, and use $\rho = -0.8, 0.5, 0.9$.
Similar to Section \ref{sec:ind_sim_reg}, the parameter of interest is $\beta$, estimator of choice is the least-squares estimate $ \hat{\beta}$, and we measure precision by constructing a 95\% confidence region for $\beta$ using the F-statistic.
We obtain the `true' value of $q_{0.95}$ from 10000 simulations.
We define error rate by $|{\hat{q}_{0.95}}/{q_{0.95}}-1|$
as before, and use block lengths $L=10,20,50$, subset sizes $b = 5000, 10000$ for BLB and SDB.
To allow BLB to complete one subset, we ran each method for 150 seconds for $L=10$, 90 seconds for $L=20$, and 60 seconds for $L=50$.

\subsection{Comparison of SDB, BLB, and MBB}
\label{sec:ts_sim_comparison}
\begin{figure}[h!]
    \centering
    \includegraphics[height=0.25\textheight]{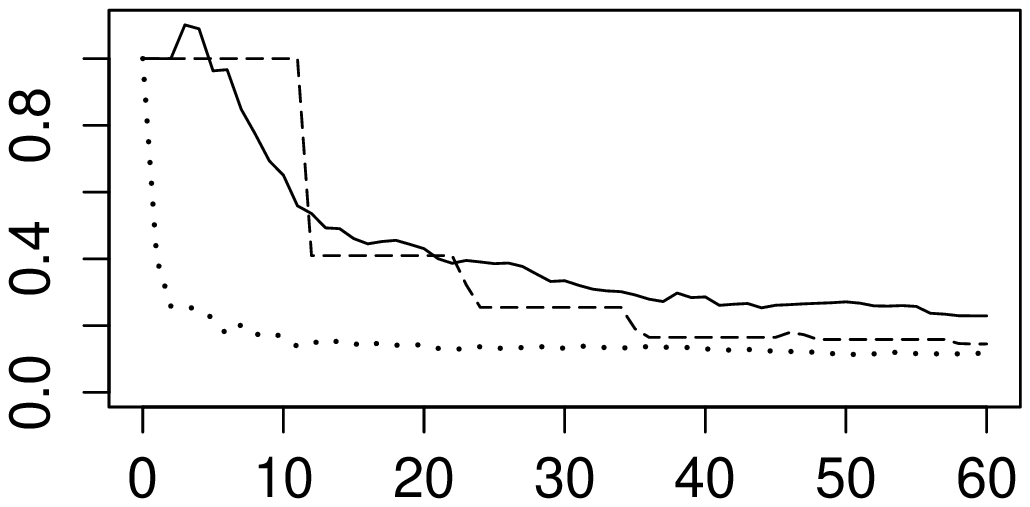}
    \includegraphics[height=0.25\textheight]{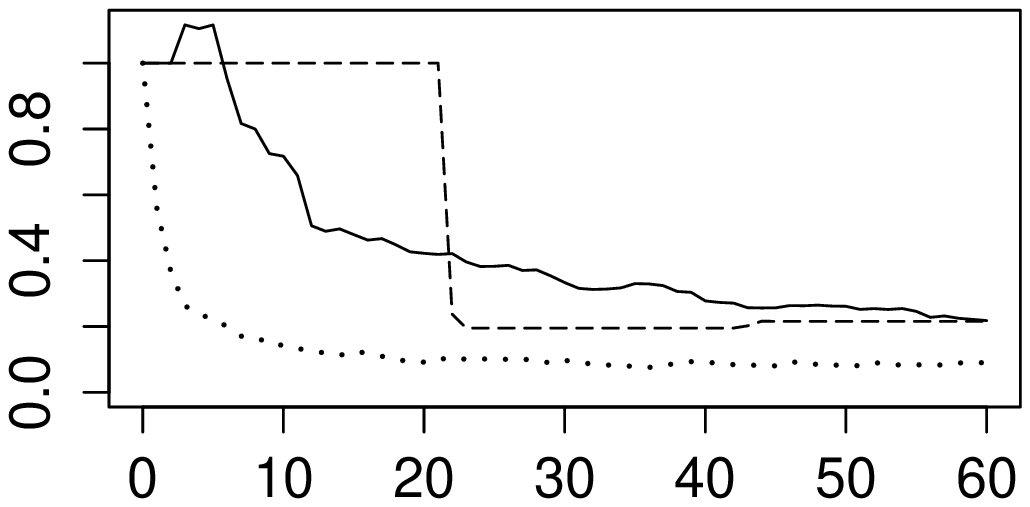}
    \includegraphics[height=0.25\textheight]{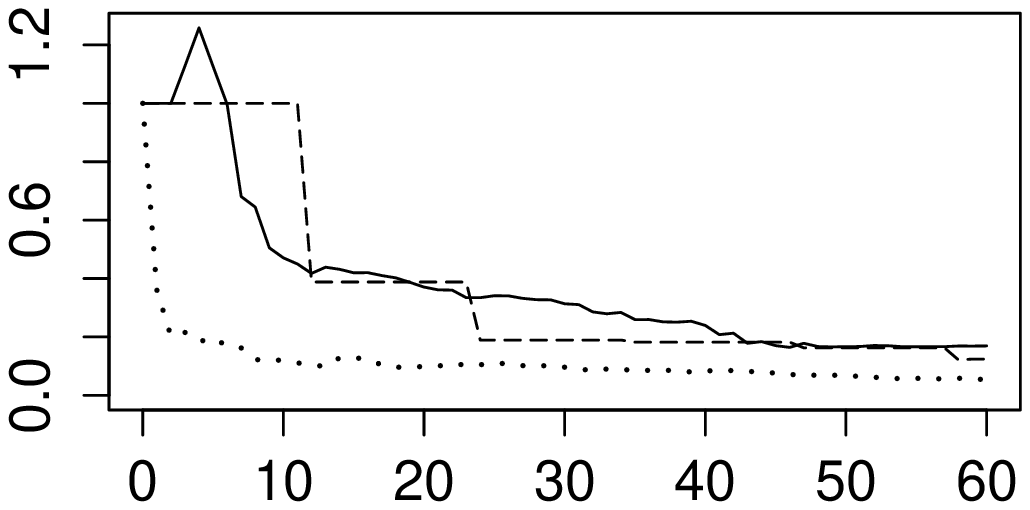}
    \includegraphics[height=0.25\textheight]{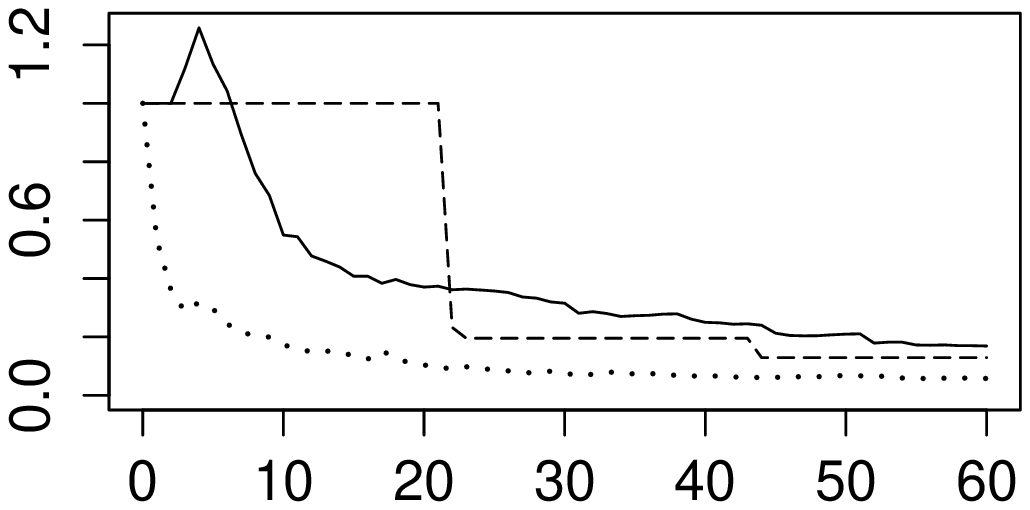}
    \includegraphics[height=0.25\textheight]{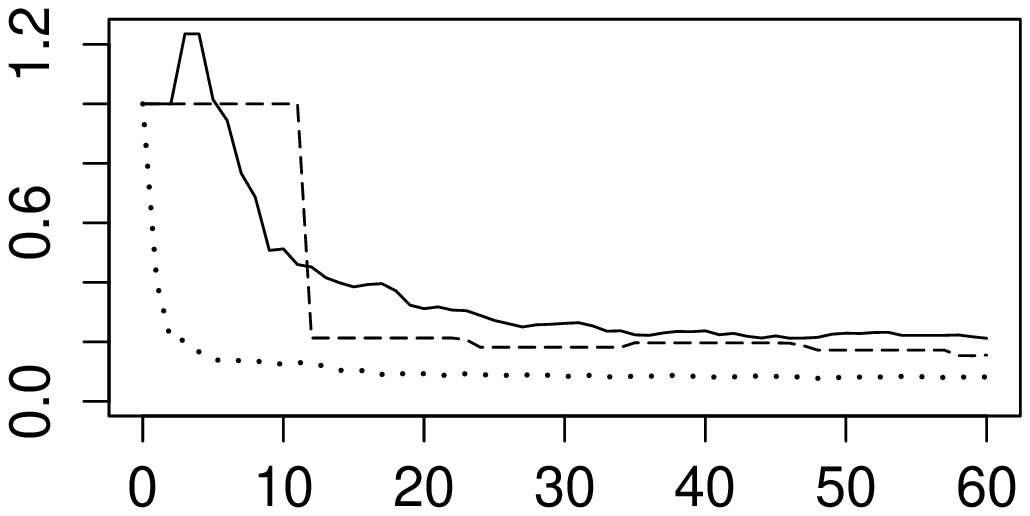}
    \includegraphics[height=0.25\textheight]{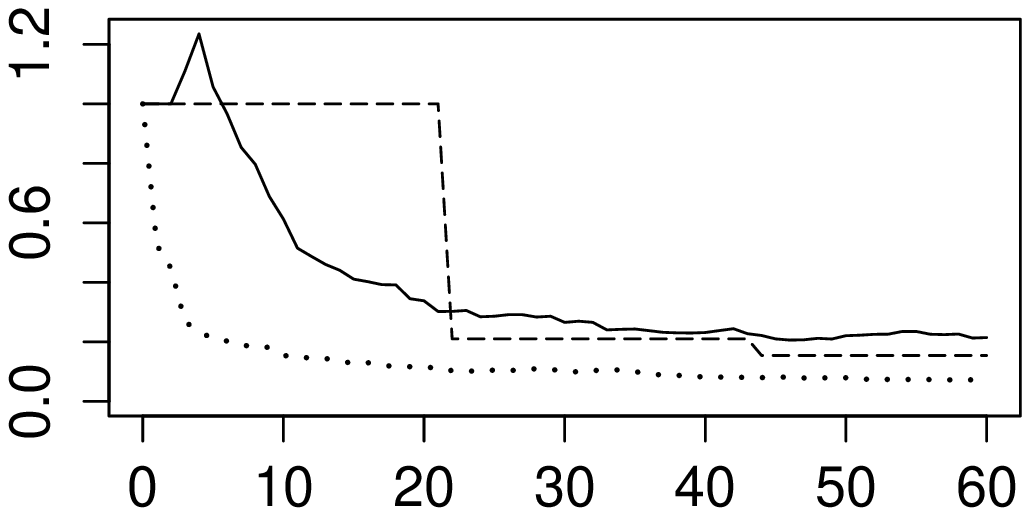}
    \caption{AR(1) simulation results with $\xi = 95\%$ quantile of $T_n = \sqrt{n}(M_n-M)$, sample size $n$=100000, block length $L$=50, autocorrelation $\rho =$ -0.8 (top row), 0.5 (middle row), 0.9 (bottom row), and subset size $b =$ 5000 (left column) ,10000 (right column).
 The plot displays the time evolution of error rates from 20 simulations when each method was allowed to run for 120 seconds.
 MBB errors are in solid lines, BLB in dashed lines, and SDB in dottted lines.}
\label{ts_Msimq1_3}
\end{figure}

The methods are compared with respect to the time evolution of error rates, as discussed in Section \ref{sec:iid_sim_comparison}.
Error rates for different methods are averaged across 20  Monte Carlo simulations.
Results for $L=50$ are displayed in Figures \ref{ts_Msimq1_3}, \ref{ts_Msimq2_3}, and \ref{ts_regsimvl3}.
To save space, results for $L=10, 20$ are presented in supplementary materials.
We can see that SDB shows significant advantages over its competitors.
In particular, it is encouraging to observe that for shorter time budgets (half of total runtime or less), SDB has a clear advantage over the other methods in most cases.
 SDB can give a reasonable estimate by 10-15 seconds in most cases, while the BLB can take a substantial time to complete a single subset. 
 MBB has highest computing cost which is reflected in its slow convergence, but it appears that  MBB can often provide a reasonable estimate by the time taken by BLB to complete one subset, which is consistent with our finding in the iid case. 

An interesting aspect of these results is that block length affects both running time and accuracy.
The behavior of the resample estimate depends on block length, and this affects accuracy of the resampling methods.
The dependence of running time on block length comes from the fact that construction of resample weights (Step (iv) of Algorithm \ref{algo-ts}) depends on the number of blocks in the resamples, and this step affects running time of the algorithms.
However, the advantages of SDB in our numerical results are consistent over the various values of block length used.


\begin{figure}[h!]
    \centering
    \includegraphics[height=0.25\textheight]{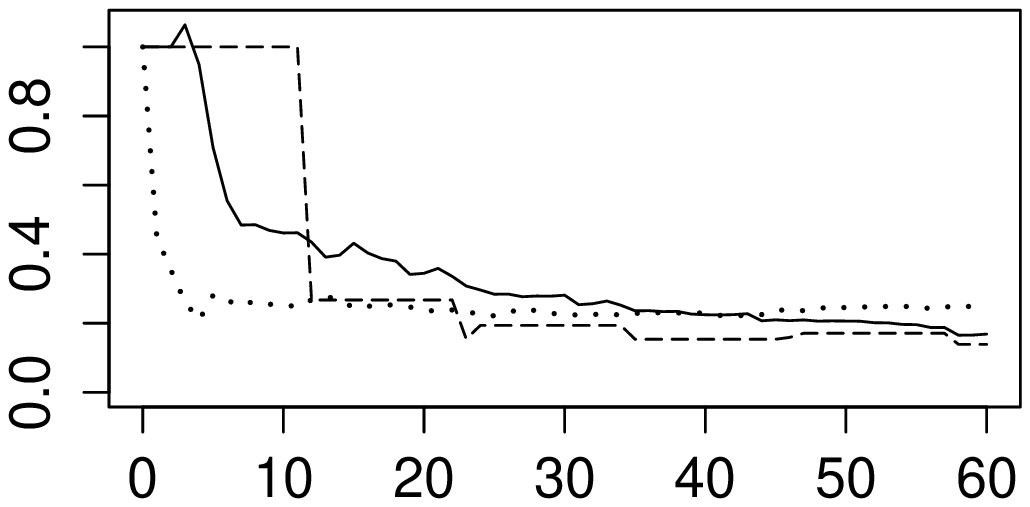}
    \includegraphics[height=0.25\textheight]{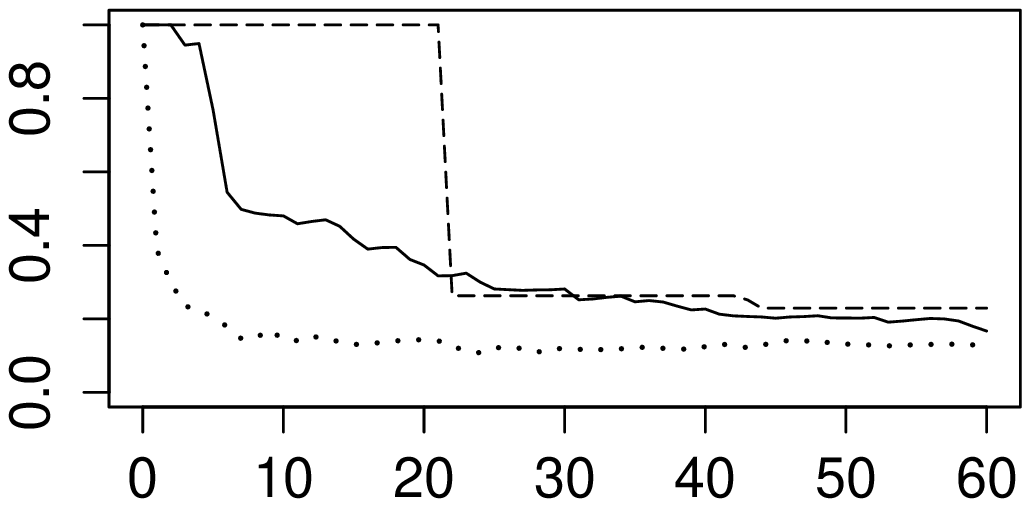}
    \includegraphics[height=0.25\textheight]{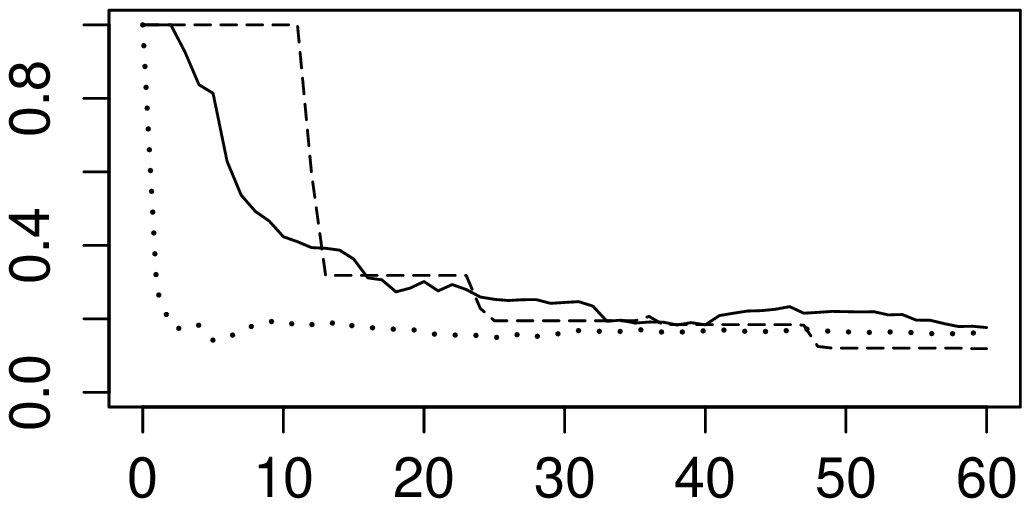}
    \includegraphics[height=0.25\textheight]{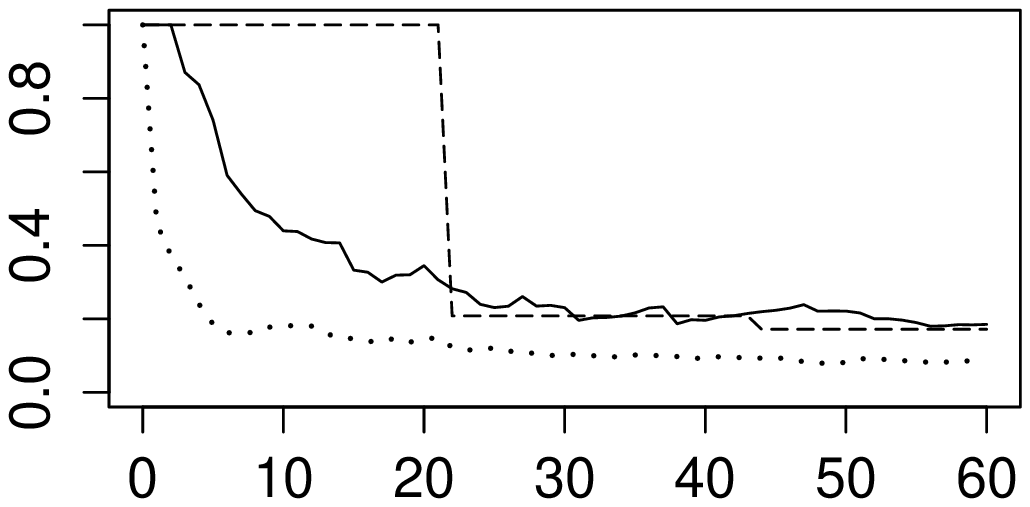}
    \includegraphics[height=0.25\textheight]{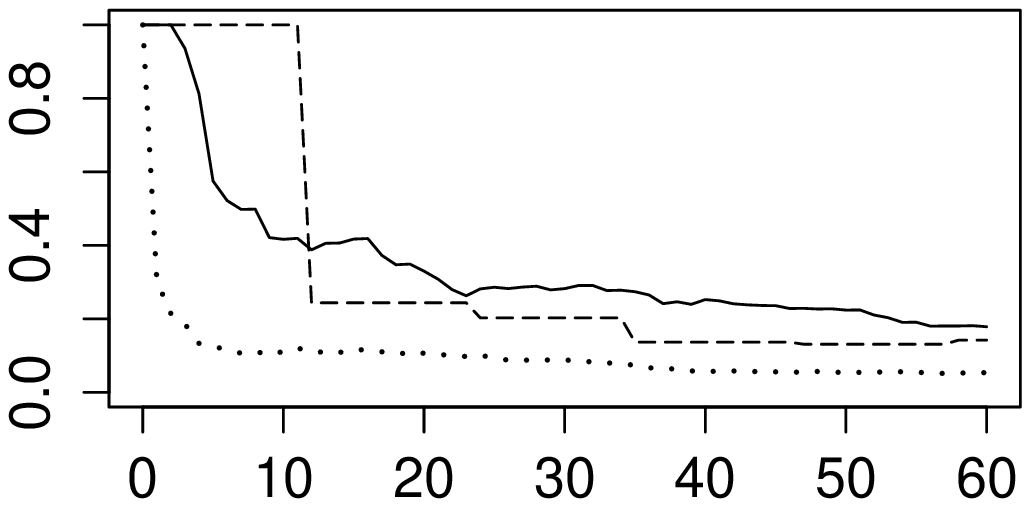}
    \includegraphics[height=0.25\textheight]{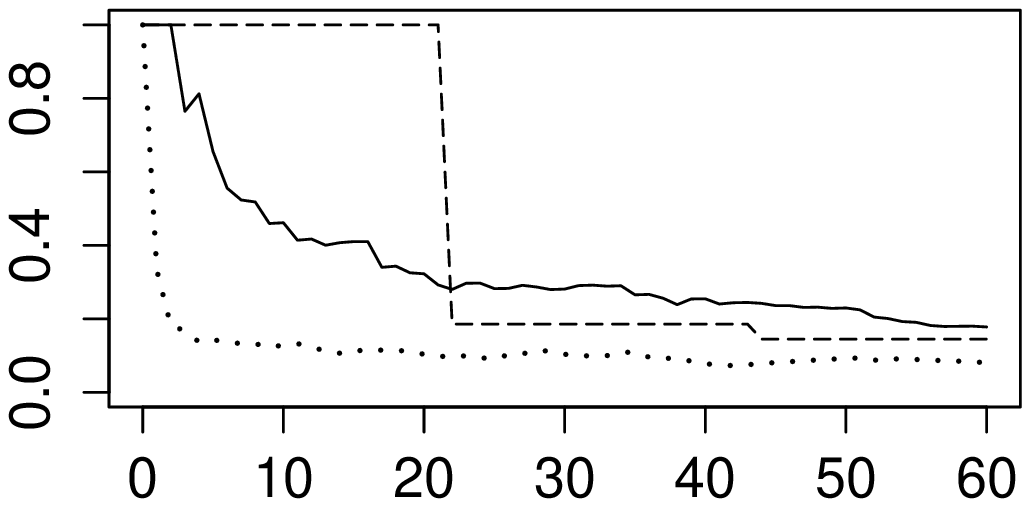}
    \caption{AR(1) simulation results with $\xi = 5\%$ quantile of $T_n = \sqrt{n}(M_n-M)$, sample size $n$=100000, block length $L$=50, autocorrelation $\rho =$ -0.8 (top row), 0.5 (middle row), 0.9 (bottom row), and subset size $b =$ 5000 (left column) ,10000 (right column).
 The plot displays the time evolution of error rates from 20 simulations when each method was allowed to run for 120 seconds.
 MBB errors are in solid lines, BLB in dashed lines, and SDB in dottted lines.}
\label{ts_Msimq2_3}
\end{figure}

\begin{figure}[h!]
    \centering
    \includegraphics[height=0.25\textheight]{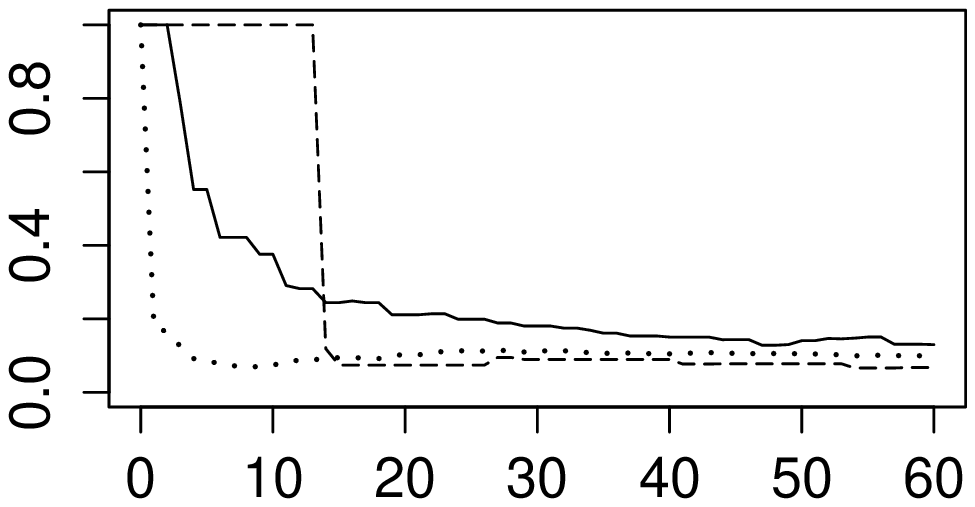}
    \includegraphics[height=0.25\textheight]{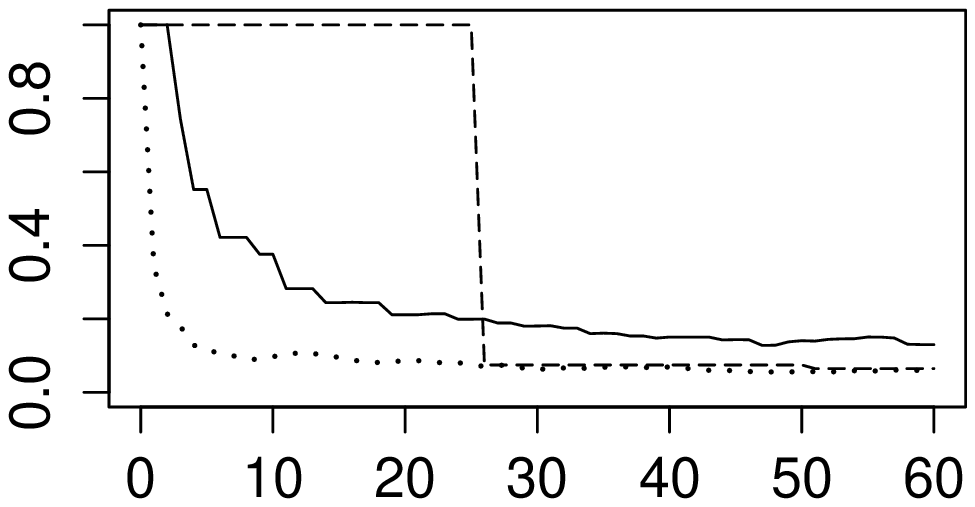}
    \includegraphics[height=0.25\textheight]{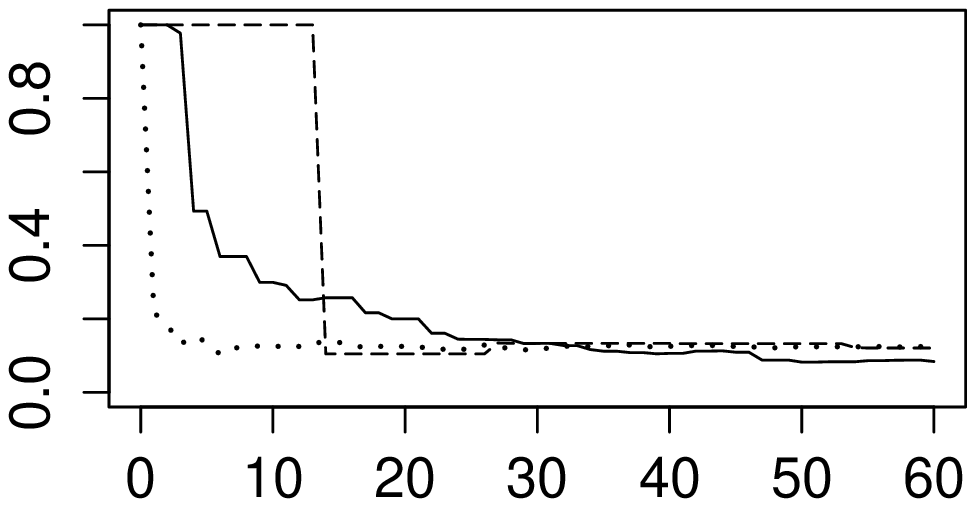}
    \includegraphics[height=0.25\textheight]{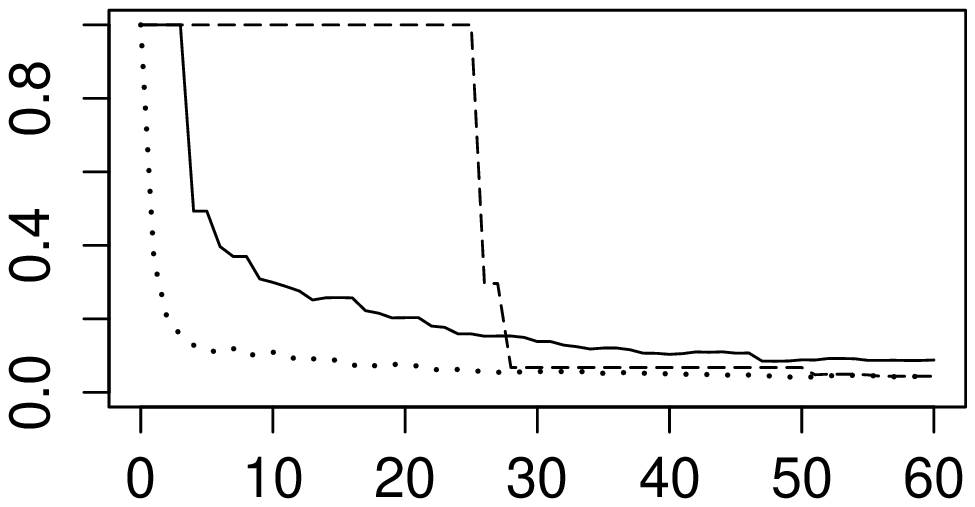}
    \includegraphics[height=0.25\textheight]{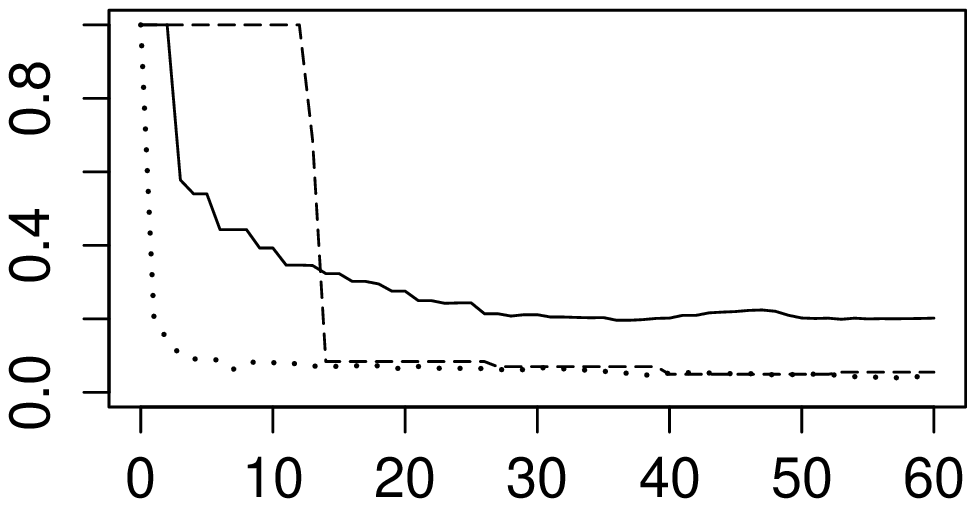}
    \includegraphics[height=0.25\textheight]{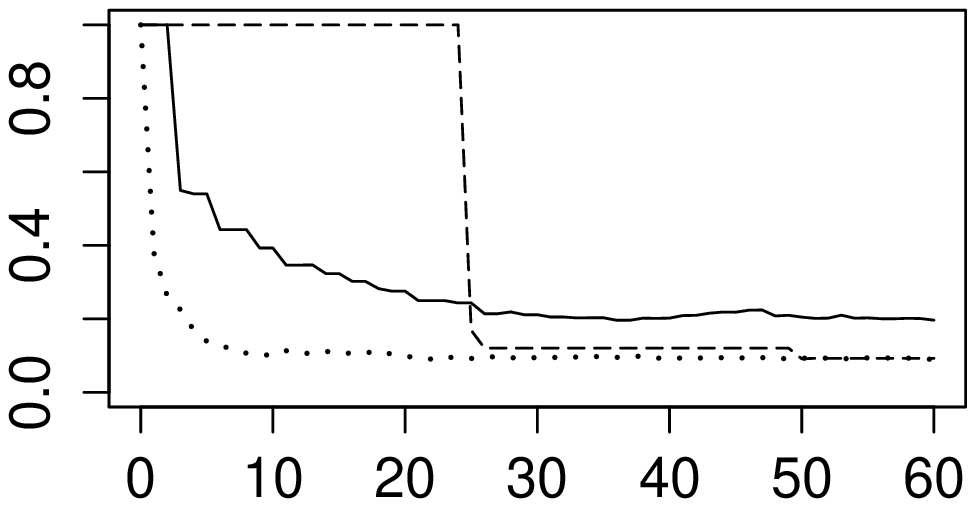}

 \caption{Time series regression simulation results with $\xi = 95\%$ quantile of $T_n = MSM/MSE$, sample size $n$=100000, dimension $d=10$,  block length $L$=50, autocorrelation $\rho =$ -0.8 (top row), 0.5 (middle row), 0.9 (bottom row), and subset size $b =$ 5000 (left column) ,10000 (right column).
 The plot displays the time evolution of error rates from 20 simulations when each method was allowed to run for 60 seconds.
 MBB errors are in solid lines, BLB in dashed lines, and SDB in dottted lines.}
\label{ts_regsimvl3}
\end{figure}


\section{Data Analysis}
\label{sec:data}
We apply our method to analyze the Central England Temperature (CET) dataset, which is a meteorological time series dataset consisting of 228 years (1780-2007) of average
daily temperatures in central England.
The CET dataset represents the longest continuous thermometer-based
temperature record on earth, and was previously analyzed by \cite{zhang2011testing} and \cite{berkes2009detecting} in the context of inference for functional time series.
In our analysis, we treat the dataset as an univariate time series sample of daily average temperatures.
The sample size is $n=228 \times 365 = 83220$, where we ignore leap years.
We remove seasonality by subtracting from each observation the mean temperature for that calendar day across $228$ years.
Our parameter of interest is the population mean $\mu$ of this univariate time series.
We use sample mean $\bar{X}$ (calculated from the $n=83,220$ observations after removing seasonality) as the estimator of $\mu$, and we want to construct a 90\% confidence interval for $\mu$ to assess the quality of estimation.
We define $T_n = \sqrt{n}(\bar{X}-\mu)$ as the root function, and let the precision measure $\xi = (q_{0.95} - q_{0.05})$ be the width of the 90\% confidence interval.

\begin{figure}[h!]
    \centering
    \includegraphics[height=0.25\textheight]{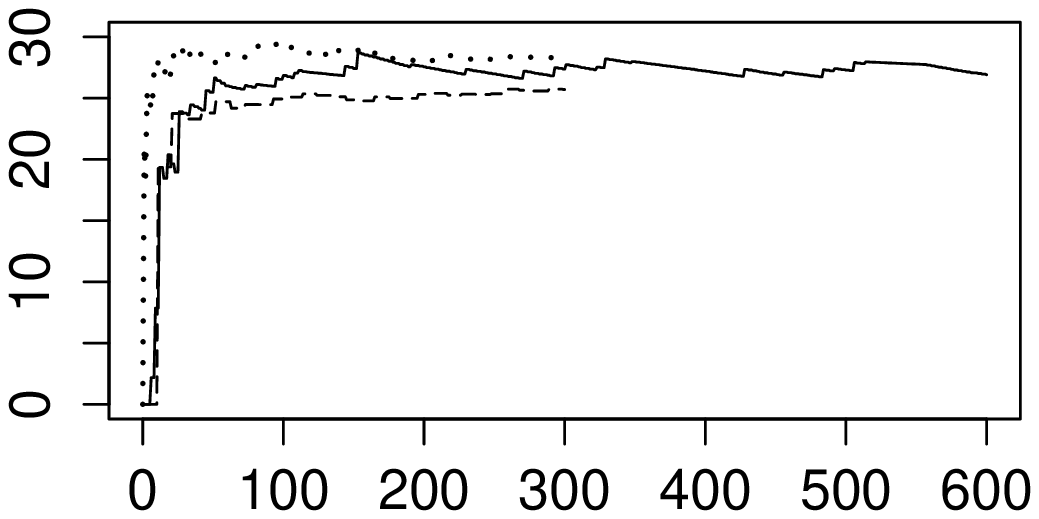}
    \includegraphics[height=0.25\textheight]{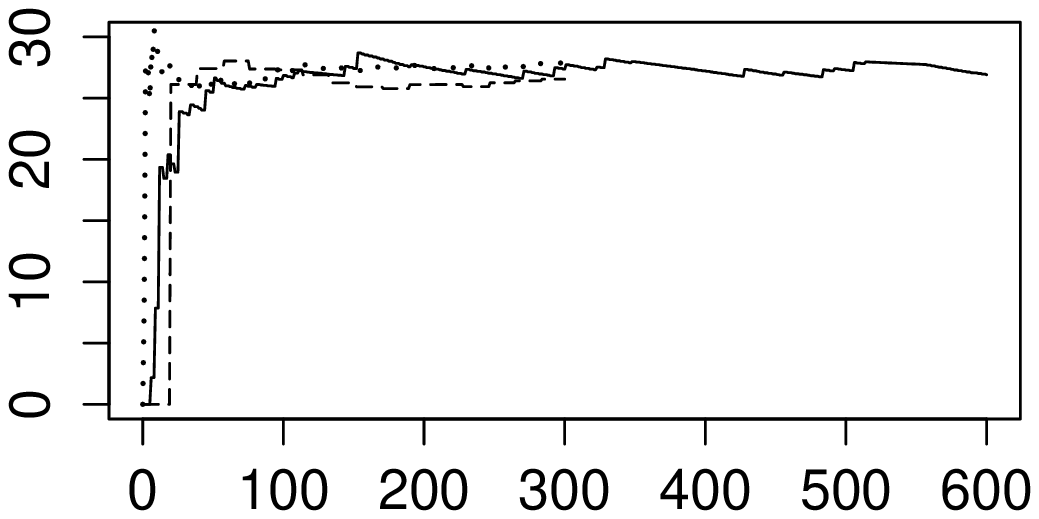}
    \includegraphics[height=0.25\textheight]{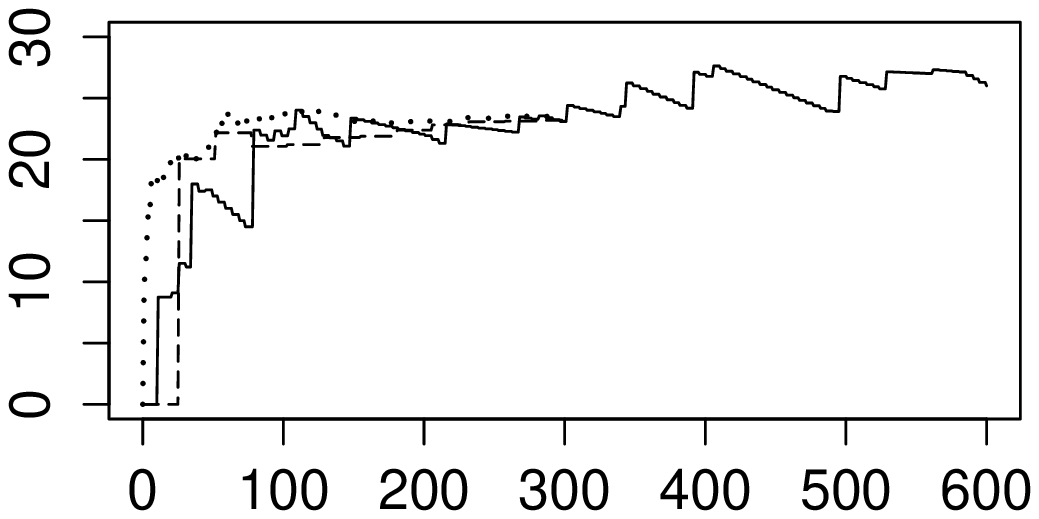}
    \includegraphics[height=0.25\textheight]{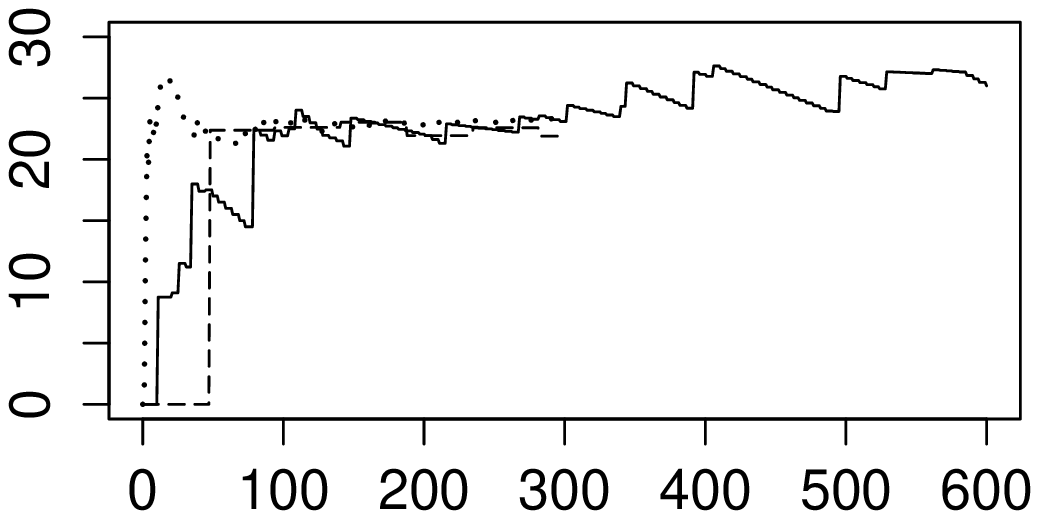}
    \includegraphics[height=0.25\textheight]{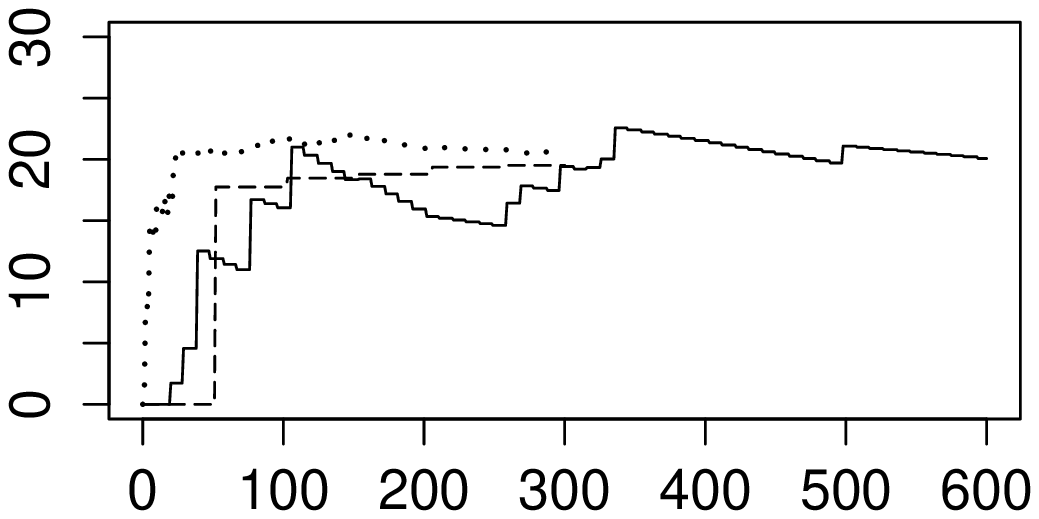}
    \includegraphics[height=0.25\textheight]{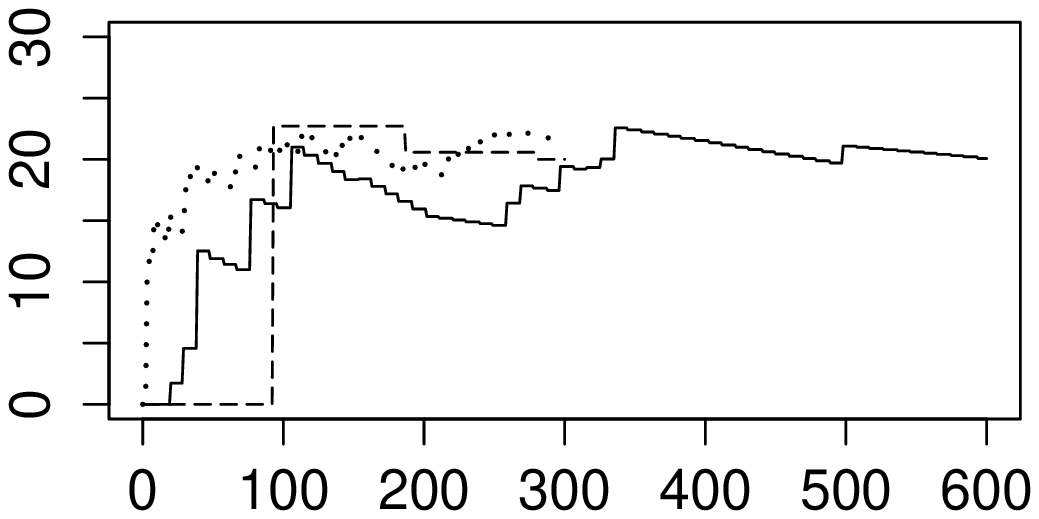}
 \caption{Time evolution of $\hat{\xi}$ for CET dataset (measured in Celsius), where $\xi = (q_{0.95} - q_{0.05})$ is the width of the 90\% confidence interval for $\mu$ based on $T_n = \sqrt{n}(\bar{X}-\mu)$. MBB was allowed to run for 600 seconds and BLB, SDB for 300 seconds. Block length $L$=50 (top row), 20 (middle row), 10 (bottom row), and subset size $b =$ 5000 (left column) ,10000 (right column).
MBB estimates are in solid lines, BLB in dashed lines, and SDB in dottted lines.}
\label{ts_CET}
\end{figure}

We applied MBB, BLB, and SDB on this dataset with block length $L=10,20,50$ and subset size $b=5000, 10000$.
MBB was allowed to run for 600 seconds, while BLB and SDB were allowed to run for 300 seconds.
Figure \ref{ts_CET} displays the time evolution of $\hat{\xi}$ for the competing methods.
Note that in this empirical example the true width is not known, however it appears that for any given block length, the three methods converge to similar estimates of the width.
MBB is the slowest to converge, and continues to display substantial oscillations well after BLB and SDB have stabilized.
BLB and SDB quickly converge to stable estimates, but for small time budgets, SDB stabilizes faster.

\section{Discussion}
\label{sec:disc}
In this article we present a new resampling method, called subsampled double bootstrap (SDB), for estimating the precision of inference methods in massive data.
Our method applies to both independent data and stationary time series data.
The main idea is to select small random subsets of the data and construct a single full size resample from each random subset, in a manner reminiscent of fast double bootstrap (\cite{white2000reality} and \cite{davidson2000improving}).
Our method inherits the theoretical strengths and automatic nature of classical resample based methods like bootstrap (\cite{efron1979bootstrap}) in the independent data context and MBB (\cite{kunsch1989jackknife}, \cite{liu1992moving}) in the time series context.
It also inherits the computational strengths of subsample based methods like subsampling (\cite{politis1994large}) and $m$ out of $n$ bootstrap (\cite{bickel1997resampling}).
The advantage of our method over the recently proposed BLB (\cite{kleiner2014scalable}) lies in sample coverage, running time, and automatic implementation without having to choose additional tuning parameters under a given time budget.
Simulation studies and data analysis examples demonstrate the advantage of our method over BLB and boostrap (i.i.d. case) or MBB (time series case) for a given computational time budget.

An important practical aspect of both SDB and BLB is the choice of subset size.
Increasing the subset size leads to increasing benefits in terms of statistical accuracy but at an increasing computational cost.
In the time series case, the regularity conditions of Theorem \ref{th:maindep} impose some restrictions on $b$ for consistency.
In practice, for a given computational time budget, it remains unclear how to choose an optimal subset size that balances statistical accuracy and running time.
A closely related problem is the selection of optimal block length for the time series version of SDB and BLB.
In the context of classical resampling methods this problem has been well studied by
\cite{hall1995blocking}, \cite{buhlmann1999block} among others.
For the time series version of SDB and BLB, we conjecture that the choice of optimal block length is associated with subset size in addition to sample size and other parameters.
We leave these interesting directions to future work. 

Further, it is worth mentioning that the higher order accuracy of BLB was studied by \cite{kleiner2014scalable} and higher order accuracy of FDB has been recently studied by \cite{chang2014double}. 
A relevant next step is a theoretical comparison of SDB and BLB which will help identify scenarios where SDB works better than BLB, or vice versa.
This comparison will involve studying higher order properties of SDB, and we plan to consider this in future research as well.


\section{Appendix: Proofs of theoretical results}
\subsection{Proof of Theorem \ref{th:SDBiid}} \label{sec:proof1}

We begin by setting up a probabilistic model for the SDB in the i.i.d. setting. When dealing with empirical processes which are defined on classes of functions, measurability questions play an important role and it is crucial to state what the underlying probability space is-- see \cite{dud1999}, Chapter 3.1 (page 91), for a discussion of related matters.
Here we will consider the following setup.

\begin{itemize}
\item[(P)] Consider a product of three probability spaces $(\Omega_i^n,\Ac_i^n,P_i^n)_{i=1,...,3}$. Assume that the observations $X_1,...,X_n$ are defined as coordinate projections on $(\Omega_1^n,\Ac_1^n,P_1^n)$, which is itself a product of $n$ identical probability spaces [this is a standard assumption in empirical process theory- see for instance \cite{dud1999}, Chapter 3.1]. Additionally, assume that on $ \Omega_2^n $ we have a random vector $(W_{1,n},...,W_{b,n}) \sim \text{Multinomial}_b(n,1/b,...,1/b)$ and that on $\Omega_3^n$ we have a random variable $R$ which follows a uniform distribution on the permutations of $\{1,...,n\}$. In what follows, denote the set of permutations of $\{1,...,n\}$ by $\sigma(n)$. Also, we assume without loss of generality that $\Omega_i^n$ are finite for $i=2,3$ and that for $i=2,3$ the sigma-algebra $\Ac_i^n$ is the power set of $\Omega_i^n$.
\end{itemize}

Throughout this proof, write $W$ for the vector $(W_{1,n},...,W_{b,n})$ and $X$ for the vector $(X_1,...,X_n)$. Define the map
\[
f\mapsto (Z_n(R,X,W))(f) := \frac{1}{\sqrt{n}}\sum_{i=1}^b (W_{i}-n/b)f(X_{R^{-1}(i)}), \quad f\in\Fc.
\]
Note that $(Z_n(R,X,W))(\cdot)$ can be viewed as an element of the space of functions $\ell^\infty(\Fc)$. Throughout the proof, denote by $\BL_1$ the set of Lipschitz continuous functions $g: \ell^\infty(\Fc)\to \R$ with Lipschitz constant $1$ that are additionally uniformly bounded by $1$. Also, use the notation $f^*,f_*$ to denote smallest measurable majorants and greatest measurable minorants, respectively. For maps of several arguments, it will sometimes be necessary to take measurable majorants and minorants with respect to only some of the arguments. For example $g(r,X,W)^{*_{X,W}}$ will be used to denote the smallest measurable majorant of the map $(x,w) \mapsto g(r,x,w)$ with $r$ being held fixed.
With this notation, we need to show that [see \cite{kos2008}, Section 2.2.3]
\begin{itemize}
\item[(i)]
\[
\sup_{h \in \BL_1} \Big|\E_{R,W} h(Z_n(R,X,W)) - \E[h(\GG)] \Big| \pstar 0.
\]
\item[(ii)] For all $h \in \BL_1$
\[
\E_{R,W} h(Z_n(R,X,W))^* - \E_{R,W} h(Z_n(R,X,W))_* \pstar 0.
\]
\end{itemize}
Here, $\E_{R,W}$ denotes the expectation with respect to $R,W$. Note that the map $(R,W)\mapsto h(Z_n(R,X,W))$ is measurable outer almost surely since $R,W$ are defined on complete, discrete probability spaces.\\
\textit{Proof of (i)}
Write
\begin{eqnarray*}
\E_{R,W} h(Z_n(R,X,W)) = \frac{1}{n!} \sum_{r \in \sigma(n)} \E_W \Big[h(Z_n(r,X,W))\Big]
\end{eqnarray*}
Then
\begin{eqnarray*}
&&\E_X^* \Big[\sup_{h \in \BL_1} \Big|\E_{R,W} h(Z_n(R,X,W)) - \E[h(\GG)]\Big| \Big]
\\
&\leq& \E_X^* \Big[\frac{1}{n!} \sum_{r \in \sigma(n)} \sup_{h \in \BL_1}  \E_W\Big| h(Z_n(r,X,W)) - \E[h(\GG)] \Big|\Big]
\\
&\leq& \E_X \Big[\frac{1}{n!} \sum_{r \in \sigma(n)}   \E_W \Big[\Big(\sup_{h \in \BL_1}\Big| h(Z_n(r,X,W)) - \E[h(\GG)] \Big|\Big)^{*_{X,W}}\Big]\Big]
\\
&=& \frac{1}{n!} \sum_{r \in \sigma(n)} \E^*\Big[\sup_{h \in \BL_1} \Big| h(Z_n(r,X,W)) - \E[h(\GG)] \Big|\Big].
\end{eqnarray*}
For each fixed value of $r$ we have
\[
(Z_n(r,X,W))(f) = \frac{1}{\sqrt{n}}\sum_{i=1}^b (W_{i,n}-n/b)f(X_{r^{-1}(i)}),
\]
which implies that $Z_n(r,X,W)$ depends on $X$ only through $(X_{r^{-1}(i)})_{i=1,...,b}$. In particular
\[
\sup_{h \in \BL_1} \Big| h(Z_n(r,x,w)) - \E[h(\GG)]\Big| = S \circ \Pi_r(x,w)
\]
where $\Pi_r(x,w) := ((x_{r^{-1}(1)},...,x_{r^{-1}(b)}),w)$ and we defined for $y,w \in \R^b$
\[
S(y,w) := \sup_{h \in \BL_1} \Big| h \Big( f \mapsto \frac{1}{\sqrt{n}}\sum_{i=1}^b (w_{i}-n/b)f(y_i) \Big) - \E[h(\GG)]\Big|.
\]
Since $X_1,...,X_n,W$ are defined on a product probability space, it follows that the measurable majorant of $S \circ \Pi_r(X,W)$ with respect to $X,W$ can be expressed as $S(\cdot,\cdot)^{*}\circ\Pi_r(x,w)$. This is a consequence of from Lemma 1.2.5 in \cite{vanwel1996} and combined with the fact that $S$ is uniformly bounded and $\Pi_r$ is a coordinate projection on a product space. In particular, the symmetry of the problem implies that
\[
\frac{1}{n!} \sum_{r \in \sigma(n)} \E^*\Big[\sup_{h \in \BL_1} \Big| h(Z_n(r,X,W)) - \E[h(\GG)] \Big|\Big] = \E^*\Big[\sup_{h \in \BL_1} \Big| h(Z_n(\text{id},X,W)) - \E[h(\GG)] \Big|\Big]
\]
where $\text{id} := (1,2,...,n)$. Moreover Theorem 3.6.3 in \cite{vanwel1996} with the identification $k_n = n, n = b$ implies that
\[
2 \geq \sup_{h \in \BL_1} \Big| h(Z_n(\text{id},X,W)) - \E[h(\GG)] \Big| \pstar 0.
\]
By dominated convergence, this yields
\[
\E^*\Big[\sup_{h \in \BL_1} \Big| h(Z_n(\text{id},X,W)) - \E[h(\GG)] \Big|\Big] \to 0,
\]
and thus statement (i) is established.\\
\\
\textit{Proof of (ii)}\\
It suffices to prove that [note that $h^* - h_* \geq 0$]
\[
\E[ h(Z_n(R,X,W))^* - h(Z_n(R,X,W))_*] \to 0.
\]
Observe that by the definition of measurable majorants we have
\begin{multline*}
h(Z_n(R,X,W))^* = \Big(\sum_{r\in\sigma(n)} I\{R=r\}  h(Z_n(r,X,W))\Big)^*
\\
\leq \sum_{r\in\sigma(n)}  I\{R=r\} \Big(h(Z_n(r,X,W))\Big)^{*_{X,W}},
\end{multline*}
since $(R,X,W) \mapsto I\{R=r\} \Big(h(Z_n(r,X,W))\Big)^{*_{X,W}}$ is measurable for each $r \in \sigma(n)$. Similarly
\begin{multline*}
h(Z_n(R,X,W))_* = \Big(\sum_{r\in\sigma(n)}  I\{R=r\}  h(Z_n(r,X,W))\Big)_*
\\
\geq \sum_{r\in\sigma(n)}  I\{R=r\} \Big(h(Z_n(r,X,W))\Big)_{*_{X,W}}.
\end{multline*}
Thus
\begin{eqnarray*}
&&\E [h(Z_n(R,X,W))^* - h(Z_n(R,X,W))_*]
\\
&\leq& \E\Big[ \sum_{r\in\sigma(n)}  I\{R=r\} \Big( \Big(h(Z_n(r,X,W))\Big)^{*_{X,W}} - \Big(h(Z_n(r,X,W))\Big)_{*_{X,W}}\Big) \Big]
\\
&=& \E\Big[\Big(h( Z_n(\text{id},X,W))\Big)^{*_{X,W}} - \Big(h(Z_n(\text{id},X,W))\Big)_{*_{X,W}}\Big]
\end{eqnarray*}
where the equality in the last line follows by arguments similar to the ones given in the proof of (i). Now the expression in the last line converges to zero by arguments similar to the ones given in the proof of (i) and Theorem 3.6.3 in \cite{vanwel1996} and thus (ii) follows. \hfill $\Box$


\subsection{Proof of Theorem \ref{th:maindep}} \label{sec:proof2} \renewcommand{\AA}{\field{A}}

Throughout this proof, we will simplify notation by assuming that $K L = n$. It is easy to see that this assumption can be relaxed.

Introduce the abbreviation $S = (s_1,...,s_K), \Xc = (X_1,...,X_n)$. Denote by $P_{S,J}$ the probability conditional on $\Xc$ and by $P_{S}$ the probability conditional on $\Xc,J$. Similarly, let $\E_{S,J}, \E_{S}, \Var_{S,J}$ and $\Var_S$ denote the corresponding versions of conditional expectations and variances. Following the discussion on page 277 in \cite{rad1996}, we will assume that all suprema we encounter are measurable. This might not always be true, but permissibility of $\Fc$ ensures that suitable modifications of our arguments remain correct [see the discussion on page 277 in \cite{rad1996}].

Define the norm $\|f\|_{p,X} := (\E[|f(X_1)|^p])^{1/p}$. For an arbitrary $\delta$-net $\Fc_\delta$ for $\Fc$ with respect to $\|\cdot\|_{p,X}$, denote by $f_\delta$ any point in $\Fc_\delta$ which minimizes $\|f-g\|_{p,X}$ over $g \in \Fc_\delta$. Consider the approximating processes $\AA_{\delta,n}^B(f) := \hat \GG_{n,b}^B(f_\delta), \AA_{\delta}(f) := \GG(f_\delta)$. By Lemma B.3 in \cite{volsha2014} the claim of the Theorem follows once we establish that
\begin{enumerate}
\item[(i)] For every $i \in \N$: $\AA_{1/i,n}^B \weakP{J,S} \AA_{1/i} $ for $n \to \infty$.
\item[(ii)] $\AA_{1/i} \weak \GG$ for $i \to \infty$.
\item[(iii)] For every $\eps > 0$: $\lim_{i \to \infty} \limsup_{n \to\infty} P(\sup_{f \in \Fc}|\AA_{1/i,n}^B(f) - \hat \GG_{n,b}^B(f)| > \eps) = 0$.
\end{enumerate}
Part (ii) follows from the properties of the limiting process $\GG$ [more precisely, there exists a version of $\GG$ with sample paths that are uniformly continuous with respect to $\|\cdot\|_{2,X}$ and thus $\|\cdot\|_{p,X}$ for $p\geq 2$ - see Theorem 2.1 in \cite{arcyu1994}]. It thus remains to establish (i) and (iii).\\

In order to establish (i), it suffices to show that for any fixed, finite collection of functions $f_1,...,f_k \in \Fc$ we have
\begin{equation} \label{eq:fidi1}
(\hat \GG_{n,b}^B(f_1),...,\hat \GG_{n,b}^B(f_k)) \weakP{J,S} (\GG(f_1),...,\GG(f_k)).
\end{equation}
Denote by $\text{BL}_1$ the set of functions on $\R^{k}$ which are bounded by $1$ and are Lipschitz continuous with Lipschitz constant bounded by $1$. In order to establish \eqref{eq:fidi1}, we need to prove that
\begin{equation} \label{eq:fidi2}
\sup_{h \in \text{BL}_1} \E_{S,J} \Big|h(\hat \GG_{n,b}^B(f_1),...,\hat \GG_{n,b}^B(f_k)) - \E[h(\GG(f_1),...,\GG(f_k))] \Big| = o_P(1)
\end{equation}
Due to the independence between $J$ and $\Xc$ and due to strict stationarity of $\{X_t\}_{t \in \Z}$, the distribution of the tuple $(X_J,...,X_{J+b-1})$ is the same as the distribution of $(X_1,...,X_{b})$ [unconditionally]. Thus the arguments on page 272 in \cite{rad1996} yield
\begin{equation} \label{eq:fidiS}
(\hat \GG_{n,b}^B(f_1),...,\hat \GG_{n,b}^B(f_k)) \weakP{S} (\GG(f_1),...,\GG(f_k)).
\end{equation}
 Observe that
\begin{multline*}
\sup_{h \in \text{BL}_1} \Big|\E_{S,J}[h(\hat \GG_{n,b}^B(f_1),...,\hat \GG_{n,b}^B(f_k))] - \E[h(\GG(f_1),...,\GG(f_k))] \Big|
\\
\leq \sup_{h \in \text{BL}_1} \E_J \E_{S} \Big|h(\hat \GG_{n,b}^B(f_1),...,\hat \GG_{n,b}^B(f_k)) -\E[h(\GG(f_1),...,\GG(f_k))] \Big|
\\
\leq \E_J \sup_{h \in \text{BL}_1} \E_{S} \Big|h(\hat \GG_{n,b}^B(f_1),...,\hat \GG_{n,b}^B(f_k)) - \E[h(\GG(f_1),...,\GG(f_k))] \Big|.
\end{multline*}
Thus for any $\eps >0$
\begin{multline*}
\E \sup_{h \in \text{BL}_1} \Big|\E_{S,J}[h(\hat \GG_{n,b}^B(f_1),...,\hat \GG_{n,b}^B(f_k))] - \E[h(\GG(f_1),...,\GG(f_k))] \Big|
\\
\leq \E\sup_{h \in \text{BL}_1} \E_{S} \Big|h(\hat \GG_{n,b}^B(f_1),...,\hat \GG_{n,b}^B(f_k)) - \E[h(\GG(f_1),...,\GG(f_k))] \Big|
\\
\leq \eps + 2 P\Big(\sup_{h \in \text{BL}_1} \E_{S} \Big|h(\hat \GG_{n,b}^B(f_1),...,\hat \GG_{n,b}^B(f_k)) - \E[h(\GG(f_1),...,\GG(f_k))] \Big| > \eps \Big)
\end{multline*}
since $\Big|h(\hat \GG_{n,b}^B(f_1),...,\hat \GG_{n,b}^B(f_k)) - \E[h(\GG(f_1),...,\GG(f_k))]  \Big| \leq 2$ by the definition of $\text{BL}_1$. By \eqref{eq:fidiS}, the probability in the last line of the above equation tends to zero [as $n \to \infty$] for any fixed $\eps > 0$, and this implies
\[
\E\sup_{h \in \text{BL}_1} \E_{S,J} \Big|h(\hat \GG_{n,b}^B(f_1),...,\hat \GG_{n,b}^B(f_k)) - \E[h(\GG(f_1),...,\GG(f_k))]  \Big| = o(1).
\]
This proves \eqref{eq:fidi2} and establishes (i).\\

{Next, let us prove (iii)}. Fix $\eps > 0$. For a function $f$, define its truncated version $f^t(x) := f(x)I\{F(x) \leq b^\gamma\}$. We shall prove that
\begin{equation}\label{eq:depboothelp1}
\sup_{f\in\Fc}| \hat \GG_{n,b}^B(f) - \hat \GG_{n,b}^B(f^t) | = o_P(1).
\end{equation}
Additionally, we will apply restricted chaining to show that there exists a sequence of sets of functions $\Fc_{n} \subset \Fc, n \in \N$ such that
\begin{multline}\label{eq:depboothelp2}
P\Big(\sup_{f,g\in \Fc, \|f-g\|_{p,X} < \delta} \Big|\hat \GG_{n,b}^B(f^t) - \hat \GG_{n,b}^B(g^t)\Big| > 3\eps \Big)
\\
\leq P\Big( \sup_{f \in \Fc_{n}g\in\Fc, \|f-g\|_{p,X}\leq (\ln b)^{-3/2}} |\hat \GG_{n,b}^B(f^t) - \hat \GG_{n,b}^B(g^t)| \geq \eps \Big) + \xi(\delta,n)
\end{multline}
where $\lim_{\delta\to 0}\lim_{n \to\infty} \xi(\delta,n) = 0$ and that
\begin{equation}\label{eq:depboothelp3}
P\Big( \sup_{f \in \Fc_{n}, g\in\Fc,\|f-g\|_{p,X}\leq (\ln b)^{-3/2}} |\hat \GG_{n,b}^B(f^t) - \hat \GG_{n,b}^B(g^t)| \geq \eps \Big) \to 0 \quad \mbox{as } n \to \infty.
\end{equation}
Taken together, \eqref{eq:depboothelp1}-\eqref{eq:depboothelp3} imply (iii). \\


Before proceeding with the proof, we remark that
\[
\hat \GG_{n,b}^B(f) = \sum_{k=1}^K  \frac{1}{\sqrt{n}}\sum_{i = 1}^L  \Big(f(X_{(k-1)L +i}^*) - \frac{1}{b} \sum_{j = 1}^{b} f(X_{J + j})\Big) =: \sum_{k=1}^K V_k(f).
\]
Note that by construction, the quantities $V_1(f),...,V_K(f)$ are independent conditionally on $J, \Xc$. Moreover, for any function $f$ which is uniformly bounded, we have that $|V_k(f)| \leq 2 n^{-1/2} L \|f\|_\infty$. Thus by Bernstein's inequality [Lemma 2.2.9 in \cite{vanwel1996}]
\begin{equation}\label{eq:bern}
P_{S}\Big( \Big|\sum_{k=1}^K V_k(f) \Big| \geq \eta \Big) \leq 2\exp\Big(-\frac{1}{2} \frac{\eta^2}{K v^2 + 2 n^{-1/2} L \|f\|_\infty\eta/3 } \Big)
\end{equation}
for any $v^2 \geq Var_S (V_1)$ [note that by construction $Var_S (V_1) = Var_S (V_k)$ for all $k$ almost surely]. Now from the definition of the bootstrap and the fact that $KL = n$
\[
K Var_S (V_k) = \frac{1}{b}\sum_{i=1}^b\Big(\frac{1}{L^{1/2}}\sum_{j=1}^L f(\tilde X_{J + i + j}) \Big)^2 - \Big(\frac{1}{b}\sum_{i=1}^b\frac{1}{L^{1/2}}\sum_{j=1}^L f(\tilde X_{J + i + j}) \Big)^2.
\]
Additionally, due to the independence between $J$ and the original sample and due to strict stationarity, the distribution of the tuple $(X_{J+1},...,X_{J+b})$ is the same as the distribution of $(X_1,...,X_{b})$ [unconditionally]. A close inspection of the proof of Lemma 3 in \cite{rad1996} [after identifying $(n,b)$ in the latter paper with $(b,L)$ in our notation] shows that under the assumption $L = o(b^\rho)$ for some $0< \rho < \frac{p-2}{2(p-1)}$ the following two claims are true:
\begin{equation} \label{eq:bootrad1}
A_n(\Gc_n) := (\ln b)^3 \sup_{h \in \Gc_n} \Big|\Var_S\Big(\frac{1}{L^{1/2}}\sum_{i=1}^L h^t(X_i^*) \Big) - \Var\Big(\frac{1}{L^{1/2}}\sum_{i=1}^L h^t(X_i) \Big)\Big| = o_P(1)
\end{equation}
for any sequence of sets $\Gc_n \subset \Fc$ with cardinality $O(n^c)$ for some fixed $c<\infty$, and
\begin{equation} \label{eq:bootrad2}
B_n := (\ln b)^2 \sup_{h \in \Fc': \|h\|_{p,X}\leq (\ln b)^{-3/2}} \Var_S\Big(\frac{1}{L^{1/2}}\sum_{i=1}^L h^t(X_i^*) \Big) = o_P(1)
\end{equation}
where $\Fc' := \{f-g: f,g\in \Fc\}$. Note that, when generating the subsamples, \cite{rad1996} uses 'wrapping' while we do not. Following the discussion in \cite{rad1996}, it is easy to see that asymptotically this does not matter.

Additionally, equation (14) in \cite{rad1996} implies that
\begin{equation} \label{eq:bootrad3}
\Var\Big(\frac{1}{L^{1/2}}\sum_{i=1}^L h^t(X_i) \Big) \leq C_0 \|h^t\|_{p,X}^2
\end{equation}
for a constant $C_0$ which depends only on $p$ and the mixing coefficients $\beta$. \\


{Proof of \eqref{eq:depboothelp1}}\\
Observe that
\[
\sup_{f\in\Fc}| \hat \GG_{n,b}^B(f) - \hat \GG_{n,b}^B(f^t)| \leq  \frac{1}{\sqrt{n}}\Big(\sum_{i=1}^n F(X_i^*)I_{\{F(X_i^*) > b^\gamma\}} + \frac{1}{b} \sum_{j = 1}^{b} F(X_{J+j})I_{\{F(X_{J+j})> b^\gamma\}}\Big).
\]
By the Chebyshev inequality it suffices to show that the expectation of the right-hand side of the above inequality is $o(1)$.
From the definition of the bootstrap, it is not difficult to see that for any $i=1,...,n$
\[
0 \leq \E [F(X_i^*)I\{F(X_i^*) > b^\gamma\}] = \E [F(X_1)I\{F(X_1) > b^\gamma\}] \leq \|F\|_{p,X} (P(F(X_1) > b^\gamma) )^{\frac{p-1}{p}}
\]
and the right-hand side of the equation above is of order $o(b^{- (p-1)\gamma})$ by dominated convergence. A similar bound holds for $\E [F(X_{J+j})I\{F(X_{J+j}) > b^\gamma\}]$. Thus \eqref{eq:depboothelp1} follows from the condition $n^{1/2} = O(b^{(p-1)\gamma})$.\\

{Proof of \eqref{eq:depboothelp2}}\\
Define $\psi_1(x) := e^x-1$ and denote by $\|\cdot\|_{\psi_1}$ the corresponding \textit{Orlitz norm} [see \cite{vanwel1996}, Chapter 2.2]. Fix $\delta > 0$ and let $k_n$ denote the smallest integer such that $\delta/2^{k_n} < (\ln b)^{-3/2}/2$. Successively construct sets $\Gc_1 \subset \Gc_2 \subset ... \subset \Gc_{k_n}$ which are maximal subsets of $\Fc$ with the property $\|f-g\|_{p,X} \geq 2^{-i} \delta$ for all $f,g \in \Gc_i$ [here, maximal means that no further element can be added to $\Gc_i$ without destroying the property that $\|f-g\|_{p,X} \geq 2^{-i} \delta$ for all $f,g \in \Gc_i$]. Observe that the cardinality of $\Gc_{k_n}$ is of polynomial order in $2^{-{k_n}}\delta$ [the cardinality of $\Gc_{k_n}$ is bounded by the packing number, which is polynomial since $\Fc$ is VC- see Theorem 2.6.7 and the discussion on page 98 in \cite{vanwel1996}], and thus of polynomial order in $n$ for any fixed $\delta$. Set $\alpha(n) := 2^{-{k_n}}\delta$ and identify the set $\Fc_{n}$ with $\Gc_{k_n}$. Next, define the event $D_n := \{A_{n}(\Fc_{n}) \leq 1\}$ and note that $P(D_n) \to 1$ for $n \to \infty$ [recall the definition of $A_{n}$ in \eqref{eq:bootrad1}] for any fixed $\delta$, this follows from \eqref{eq:bootrad1}. Observe that $I_{D_n}$ is independent of $S$ and that by definition of $D_n$ we have for any $f \in \Fc_{\alpha(n)}$ and any $\eta >0$
\begin{multline*}
 P\Big( |\hat \GG_{n,b}^B(f^t)|I_{D_n} > \eta \Big) = \E \E_S I_{\{|\hat \GG_{n,b}^B(f^t)| > \eta\}} I_{D_n}
\\
 \leq 2 \E\Big[I_{D_n} \exp\Big(-\frac{1}{2} \frac{\eta^2}{\Var_S\Big(L^{-1/2}\sum_{i=1}^L h(X_i^*) \Big) + \frac{2}{3} n^{-1/2} L b^\gamma \eta} \Big) \Big]
\\
 \leq 2 \E\Big[I_{D_n} \exp\Big(-\frac{1}{2} \frac{\eta^2}{C_0 \|f^t\|_{p,X}^2 + (\ln b)^{-3} + \frac{2}{3} n^{-1/2} L b^\gamma \eta} \Big) \Big]
\\
 \leq 2 \exp\Big(-\frac{1}{2} \frac{\eta^2}{C_0 \|f^t\|_{p,X}^2 + (\ln b)^{-3} + \frac{2}{3} n^{-1/2} L b^\gamma \eta} \Big)
\end{multline*}
where the first inequality follows from \eqref{eq:bern} and the second from the definition of $D_n$. From the inequality above combined with Lemma 2.2.10 in \cite{vanwel1996} [applied with $m=1$] we obtain that for any $f \in \Fc_{n}$
\[
\Big\|\hat \GG_{n,b}^B(f^t)I_{D_n}\Big\|_{\psi_1} \leq C\Big(n^{-1/2} L b^\gamma + (\|f^t\|_{p,X}^2 + (\ln b)^{-3} )^{1/2}\Big)
\]
for some constant $C$ that does not depend on $f, \delta, n$. In particular we obtain for sufficiently large $n$ [due to the assumptions on $L,b,n$]
\begin{equation} \label{eq:kley1}
\Big\|\hat \GG_{n,b}^B(f^t)I_{D_n}\Big\|_{\psi_1} \leq C'\|f^t\|_{p,X} \quad \forall f \in \Fc_{n}: \|f^t\|_{p,X} \geq (\ln b)^{-3/2}/2.
\end{equation}
We now shall apply Lemma 7.1 from \cite{klevoldethal2014}. In the notation of that Lemma, let $T := \Fc, d(f,g) := \|f^t-g^t\|_{p,X}, \bar \eta := (\ln b)^{-3/2}, \Psi := \psi_1, \eta = \delta, \GG_f := \hat \GG_{n,b}^B(f^t)I_{D_n}$. A careful inspection of the proof of that Lemma reveals that \eqref{eq:kley1} is already sufficient to obtain the bound
\begin{multline*}
\sup_{f,g\in \Fc: \|f-g\|_{p,X} \leq \delta} |\hat \GG_{n,b}^B(f^t) - \hat \GG_{n,b}^B(g^t)|I_{D_n}
\\
\leq S_1 + 2 \sup_{f \in \Fc_{n},g\in\Fc, \|f-g\|_{p,X}\leq (\ln b)^{-3/2}} |\hat \GG_{n,b}^B(f^t) - \hat \GG_{n,b}^B(g^t)|I_{D_n}
\end{multline*}
where $S_1$ is such that [note that $\psi_1^{-1}(x) = \ln(1+x)$ and that the packing number of $\Fc$ with respect to $\|\cdot\|_{p,X}$ is of polynomial order since $\Fc$ is VC- see Theorem 2.6.7 and the discussion on page 98 in \cite{vanwel1996}]
\[
\|S_1\|_{\psi_1} \leq C \Big[\int_{(\ln b)^{-3/2}/2}^\delta 1 + |\log \eps| d\eps + (\delta + 2(\ln b)^{-3/2})(1 + |\log \delta|) \Big]
\]
for some constant $C$ independent of $\delta,n$. To complete the proof of \eqref{eq:depboothelp2}, observe that
\begin{align*}
& P\Big(\sup_{f,g\in \Fc, \|f-g\|_{p,X} < \delta} \Big|\hat \GG_{n,b}^B(f^t) - \hat \GG_{n,b}^B(g^t)\Big| > 3\eps \Big)
\\
&  \leq P\Big(\sup_{f,g\in \Fc, \|f-g\|_{p,X} < \delta} \Big|\hat \GG_{n,b}^B(f^t) - \hat \GG_{n,b}^B(g^t)\Big|I_{D_n} > 3\eps \Big) + 1 - P(D_n)
\\
& \leq P(|S_1| > \eps) + P\Big( \sup_{f \in \Fc_{n},g\in\Fc, \|f-g\|_{p,X}\leq (\ln b)^{-3/2}} |\hat \GG_{n,b}^B(f^t) - \hat \GG_{n,b}^B(g^t)| \geq \eps \Big)
\\
& + 1 - P(D_n).
\end{align*}
Setting $\xi(n,\delta) := P(|S_1| > \eps) + 1 - P(D_n)$ completes the proof of \eqref{eq:depboothelp2}.\\


{Proof of \eqref{eq:depboothelp3}}\\
Define $P_n f := \frac{1}{n}\sum_{i=1}^n f(X_{i})$ and consider the pseudo-distance $d_n(f,g) := P_n|f - g|$. Observe that to each $f \in \Fc$ we can attach a $\tilde f \in \Fc_{n}$ such that $\|f-\tilde f\|_{p,X} \leq (\ln b)^{-3/2}$. Let $\Hc := \{f^t-\tilde f^t: f \in \Fc\}$ and denote by $\Hc_n$ an $n^{-2}$ net for $\Hc$ under $d_n$. To each $h \in \Hc$, attach a $\tilde h \in \Hc$ such that $d_n(h,\tilde h) \leq n^{-2}$. Since $\Fc$ is VC, $\Hc_n$ can be chosen such that, for $n$ sufficiently large, the cardinality of $\Hc_n$ is bounded by $C (P_n F)^c n^c$ for some fixed constants $C,c$ which do not depend on $J,S$. Define the event [recall the definition of $B_n$ in \eqref{eq:bootrad3}]
\[
\tilde D_n := \Big\{B_{n} \vee \Big|{n}^{-1} \sum_{i=1}^{n} F(X_i) - \E[F(X_1)] \Big| \leq 1 \Big\}
\]
and note that by \eqref{eq:bootrad3} $P(\tilde D_n) \to 1$ [since under the assumptions of the present Theorem $P_n F - \E F[X_1] \to 0$ in probability] and that $I_{\tilde D_n}$ is independent of $S$. Observe that
\[
\sup_{f \in \Fc_{n}, \|f-g\|_{p,X}\leq (\ln b)^{-3/2}} |\hat \GG_{n,b}^B(f^t) - \hat \GG_{n,b}^B(g^t)|
\leq \sup_{h\in \Hc} \Big|\hat \GG_{n,b}^B(\tilde h)\Big| + \sup_{h\in \Hc} \Big|\hat \GG_{n,b}^B(h- \tilde h)\Big|
\]
and that for any function $f$
\begin{align*}
\Big|\hat \GG_{n,b}^B(f)\Big| & \leq \frac{1}{\sqrt{n}} \sum_{i=1}^n |f(X_i^*)| + \frac{\sqrt{n}}{b} \sum_{i=0}^{b-1} |f(X_{J+i})| \leq \frac{n^2}{\sqrt{n}} \frac{1}{n}\sum_{i=1}^n |f(X_i)| + \frac{n\sqrt{n}}{b} \frac{1}{n}\sum_{i=1}^{n} |f(X_{i})|
\\
& \leq 2n^{3/2} \frac{1}{n}\sum_{i=1}^{n} |f(X_{i})|.
\end{align*}
Thus by definition of $\tilde h$
\[
\sup_{h\in \Hc} \Big|\hat \GG_{n,b}^B(h - \tilde h)\Big| \leq 2 n^{3/2} \sup_{h\in \Hc} \frac{1}{n}\sum_{i=1}^{n} |(h-\tilde h)(X_{i})| \leq 2n^{-1/2}.
\]
Hence it suffices to show $\sup_{h\in \Hc} \Big|\hat \GG_{n,b}^B(\tilde h)\Big| = o_P(1).$ To this end, note that on $\tilde D_n$ the cardinality of $\Hc_n$ is bounded by $C'n^{c'}$ for some constants $C',c'$ independent of $n$. Thus
\begin{align*}
& P\Big(\sup_{h\in \Hc} \Big|\hat \GG_{n,b}^B(\tilde h)\Big| > \tau \Big)
\\
& \leq P\Big( \Big\{\sup_{h\in \Hc_n} \Big|\hat \GG_{n,b}^B(h)\Big| > \tau\Big\} \cap \tilde D_n \Big) + 1 - P(\tilde D_n)
\\
& = \E \E_S \Big[ I_{\{\sup_{h\in \Hc_n} |\hat \GG_{n,b}^B(h)| > \tau\}}I_{\tilde D_n} \Big]+ o(1)
\\
& \leq \E \E_S \Big[ \sum_{h \in \Hc_n}I_{\{|\hat \GG_{n,b}^B(h)| > \tau\}}I_{\tilde D_n} \Big]+ o(1)
\\
& = \E  \Big[ I_{\tilde D_n} \sum_{h \in \Hc_n}\E_S[I_{\{|\hat \GG_{n,b}^B(h)| > \tau\}}]\Big]+ o(1)
\\
& = C'n^{c'}  \E \Big[\sup_{h\in \Hc}I_{\tilde D_n} P_S (|\hat \GG_{n,b}^B(h)| > \tau )\Big] + o(1)
\\
& \text{by \eqref{eq:bern}}
\\
& \leq 2 C'n^{c'} \E \Big[\sup_{h\in \Hc} \exp\Big(-\frac{1}{2} \frac{\tau^2}{\Var_S\Big(L^{-1/2}\sum_{i=1}^L h(X_i^*) \Big) + \frac{4}{3} n^{-1/2} L b^\gamma \tau} \Big)I_{\tilde D_n}\Big] + o(1)
\\
& \leq 2 C'n^{c'}  \E\Big[ \exp\Big(-\frac{1}{2} \frac{\tau^2}{\sup_{h\in \Hc}\Var_S\Big(L^{-1/2}\sum_{i=1}^L h(X_i^*) \Big) + \frac{4}{3} n^{-1/2} L b^\gamma \tau} \Big)I_{\tilde D_n}\Big] + o(1)
\\
&\text{by the definition of $\Hc$ and $\tilde D_n$}
\\
& \leq 2 C'n^{c'}  \E\Big[ \exp\Big(-\frac{1}{2} \frac{\tau^2}{(\ln b)^{-2} + \frac{4}{3} n^{-1/2} L b^\gamma \tau} \Big)I_{\tilde D_n}\Big] + o(1)
\\
& \leq 2 C'n^{c'} \exp\Big(-\frac{1}{2} \frac{\tau^2}{(\ln b)^{-2} + \frac{4}{3} n^{-1/2} L b^\gamma \tau} \Big) + o(1)
\\
& \text{since $(\ln n)^2 = o(n^{-1/2} L b^\gamma)$ by the assumptions on $L,b,n$}
\\
& = o(1).
\end{align*}
This shows that $\sup_{h\in \Hc} \Big|\hat \GG_{n,b}^B(\tilde h)\Big| = o_P(1)$ and completes the proof of \eqref{eq:depboothelp3}. \hfill $\Box$

\bibliographystyle{chicago}
\bibliography{ref}

\begin{thebibliography}{}

\bibitem[\protect\citeauthoryear{Ahlgren and Antell}{Ahlgren and
  Antell}{2008}]{ahlgren2008bootstrap}
Ahlgren, N. and J.~Antell (2008).
\newblock Bootstrap and fast double bootstrap tests of cointegration rank with
  financial time series.
\newblock {\em Computational Statistics and Data Analysis\/}~{\em 52\/}(10),
  4754--4767.

\bibitem[\protect\citeauthoryear{Andrews and Monahan}{Andrews and
  Monahan}{1992}]{andrews1992improved}
Andrews, D.~W. and J.~C. Monahan (1992).
\newblock An improved heteroskedasticity and autocorrelation consistent
  covariance matrix estimator.
\newblock {\em Econometrica\/}~{\em 60\/}(4), 953--966.

\bibitem[\protect\citeauthoryear{Arcones and Yu}{Arcones and
  Yu}{1994}]{arcyu1994}
Arcones, M.~A. and B.~Yu (1994).
\newblock Central limit theorems for empirical andu-processes of stationary
  mixing sequences.
\newblock {\em Journal of Theoretical Probability\/}~{\em 7\/}(1), 47--71.

\bibitem[\protect\citeauthoryear{Beran}{Beran}{1988}]{beran1988prepivoting}
Beran, R. (1988).
\newblock Prepivoting test statistics: a bootstrap view of asymptotic
  refinements.
\newblock {\em Journal of the American Statistical Association\/}~{\em
  83\/}(403), 687--697.

\bibitem[\protect\citeauthoryear{Berkes, Gabrys, Horv{\'a}th, and
  Kokoszka}{Berkes et~al.}{2009}]{berkes2009detecting}
Berkes, I., R.~Gabrys, L.~Horv{\'a}th, and P.~Kokoszka (2009).
\newblock Detecting changes in the mean of functional observations.
\newblock {\em Journal of the Royal Statistical Society: Series B (Statistical
  Methodology)\/}~{\em 71\/}(5), 927--946.

\bibitem[\protect\citeauthoryear{Bickel, G{\"o}tze, and van Zwet}{Bickel
  et~al.}{1997}]{bickel1997resampling}
Bickel, P., F.~G{\"o}tze, and W.~van Zwet (1997).
\newblock Resampling fewer than $n$ observations: Gains, losses, and remedies
  for losses.
\newblock {\em Statistica Sinica\/}~{\em 7}, 1--31.

\bibitem[\protect\citeauthoryear{B{\"u}hlmann}{B{\"u}hlmann}{1995}]{buh1995}
B{\"u}hlmann, P. (1995).
\newblock The blockwise bootstrap for general empirical processes of stationary
  sequences.
\newblock {\em Stochastic Processes and their Applications\/}~{\em 58\/}(2),
  247--265.

\bibitem[\protect\citeauthoryear{B{\"u}hlmann and K{\"u}nsch}{B{\"u}hlmann and
  K{\"u}nsch}{1999}]{buhlmann1999block}
B{\"u}hlmann, P. and H.~R. K{\"u}nsch (1999).
\newblock Block length selection in the bootstrap for time series.
\newblock {\em Computational Statistics and Data Analysis\/}~{\em 31\/}(3),
  295--310.

\bibitem[\protect\citeauthoryear{Chang and Hall}{Chang and
  Hall}{2014}]{chang2014double}
Chang, J. and P.~Hall (2014).
\newblock Double-bootstrap methods that use a single double-bootstrap
  simulation.
\newblock {\em arXiv preprint arXiv:1408.6327\/}.

\bibitem[\protect\citeauthoryear{Davidson and MacKinnon}{Davidson and
  MacKinnon}{2000}]{davidson2000improving}
Davidson, R. and J.~G. MacKinnon (2000).
\newblock Improving the reliability of bootstrap tests.
\newblock Queen’s University Working paper no. 995.

\bibitem[\protect\citeauthoryear{Davidson and MacKinnon}{Davidson and
  MacKinnon}{2002}]{davidson2002fast}
Davidson, R. and J.~G. MacKinnon (2002).
\newblock Fast double bootstrap tests of nonnested linear regression models.
\newblock {\em Econometric Reviews\/}~{\em 21\/}(4), 419--429.

\bibitem[\protect\citeauthoryear{Davidson and MacKinnon}{Davidson and
  MacKinnon}{2007}]{davidson2007improving}
Davidson, R. and J.~G. MacKinnon (2007).
\newblock Improving the reliability of bootstrap tests with the fast double
  bootstrap.
\newblock {\em Computational Statistics and Data Analysis\/}~{\em 51\/}(7),
  3259--3281.

\bibitem[\protect\citeauthoryear{Dudley}{Dudley}{1999}]{dud1999}
Dudley, R.~M. (1999).
\newblock {\em Uniform Central Limit Theorems}, Volume~23.
\newblock Cambridge University Press.

\bibitem[\protect\citeauthoryear{Efron}{Efron}{1979}]{efron1979bootstrap}
Efron, B. (1979).
\newblock Bootstrap methods: Another look at the jackknife.
\newblock {\em The Annals of Statistics\/}~{\em 7\/}(1), 1--26.

\bibitem[\protect\citeauthoryear{Giacomini, Politis, and White}{Giacomini
  et~al.}{2013}]{giacomini2013warp}
Giacomini, R., D.~N. Politis, and H.~White (2013).
\newblock A warp-speed method for conducting monte carlo experiments involving
  bootstrap estimators.
\newblock {\em Econometric Theory\/}~{\em 29\/}(3), 567--589.

\bibitem[\protect\citeauthoryear{Gin{\'e} and Zinn}{Gin{\'e} and
  Zinn}{1984}]{ginzin1984}
Gin{\'e}, E. and J.~Zinn (1984).
\newblock Some limit theorems for empirical processes.
\newblock {\em The Annals of Probability\/}~{\em 12\/}(4), 929--989.

\bibitem[\protect\citeauthoryear{Hall, Horowitz, and Jing}{Hall
  et~al.}{1995}]{hall1995blocking}
Hall, P., J.~L. Horowitz, and B.-Y. Jing (1995).
\newblock On blocking rules for the bootstrap with dependent data.
\newblock {\em Biometrika\/}~{\em 82\/}(3), 561--574.

\bibitem[\protect\citeauthoryear{Jordan}{Jordan}{2013}]{jordan2013statistics}
Jordan, M.~I. (2013).
\newblock On statistics, computation and scalability.
\newblock {\em Bernoulli\/}~{\em 19\/}(4), 1378--1390.

\bibitem[\protect\citeauthoryear{Kiefer, Vogelsang, and Bunzel}{Kiefer
  et~al.}{2000}]{kiefer2000simple}
Kiefer, N.~M., T.~J. Vogelsang, and H.~Bunzel (2000).
\newblock Simple robust testing of regression hypotheses.
\newblock {\em Econometrica\/}~{\em 68\/}(3), 695--714.

\bibitem[\protect\citeauthoryear{Kleiner, Talwalkar, Sarkar, and
  Jordan}{Kleiner et~al.}{2014}]{kleiner2014scalable}
Kleiner, A., A.~Talwalkar, P.~Sarkar, and M.~I. Jordan (2014).
\newblock A scalable bootstrap for massive data.
\newblock {\em Journal of the Royal Statistical Society: Series B (Statistical
  Methodology)\/}~{\em 76}, 795--–816.

\bibitem[\protect\citeauthoryear{Kley, Volgushev, Dette, and Hallin}{Kley
  et~al.}{2014}]{klevoldethal2014}
Kley, T., S.~Volgushev, H.~Dette, and M.~Hallin (2014).
\newblock Quantile spectral processes: Asymptotic analysis and inference.
\newblock {\em arXiv preprint arXiv:1401.8104\/}.

\bibitem[\protect\citeauthoryear{Kosorok}{Kosorok}{2008}]{kos2008}
Kosorok, M.~R. (2008).
\newblock {\em Introduction to Empirical Processes and Semiparametric
  Inference}.
\newblock New York: Springer.

\bibitem[\protect\citeauthoryear{K{\"u}nsch}{K{\"u}nsch}{1989}]{kunsch1989jackknife}
K{\"u}nsch, H.~R. (1989).
\newblock The jackknife and the bootstrap for general stationary observations.
\newblock {\em The Annals of Statistics\/}~{\em 17\/}(3), 1217--1241.

\bibitem[\protect\citeauthoryear{Laptev, Zaniolo, and Lu}{Laptev
  et~al.}{2012}]{laptevboot}
Laptev, N., C.~Zaniolo, and T.-C. Lu (2012).
\newblock {BOOT-TS: A Scalable Bootstrap for Massive Time-Series Data}.
\newblock
  \url{http://cs.ucla.edu/~zaniolo/papers/biglearning2012_submission_3.pdf}.
\newblock [Online; accessed 13-December-2014].

\bibitem[\protect\citeauthoryear{Liu and Singh}{Liu and
  Singh}{1992}]{liu1992moving}
Liu, R.~Y. and K.~Singh (1992).
\newblock Moving blocks jackknife and bootstrap capture weak dependence.
\newblock {\em Exploring the limits of bootstrap\/}, 225--248.

\bibitem[\protect\citeauthoryear{Politis and Romano}{Politis and
  Romano}{1994a}]{politis1994large}
Politis, D.~N. and J.~P. Romano (1994a).
\newblock Large sample confidence regions based on subsamples under minimal
  assumptions.
\newblock {\em The Annals of Statistics\/}~{\em 22\/}(4), 2031--2050.

\bibitem[\protect\citeauthoryear{Politis and Romano}{Politis and
  Romano}{1994b}]{politis1994stationary}
Politis, D.~N. and J.~P. Romano (1994b).
\newblock The stationary bootstrap.
\newblock {\em Journal of the American Statistical Association\/}~{\em
  89\/}(428), 1303--1313.

\bibitem[\protect\citeauthoryear{Radulovi{\'c}}{Radulovi{\'c}}{1996}]{rad1996}
Radulovi{\'c}, D. (1996).
\newblock The bootstrap for empirical processes based on stationary
  observations.
\newblock {\em Stochastic processes and their applications\/}~{\em 65\/}(2),
  259--279.

\bibitem[\protect\citeauthoryear{Radulovi{\'c}}{Radulovi{\'c}}{2002}]{rad2002}
Radulovi{\'c}, D. (2002).
\newblock On the bootstrap and empirical processes for dependent sequences.
\newblock In {\em Empirical process techniques for dependent data}, pp.\
  345--364. Springer.

\bibitem[\protect\citeauthoryear{Radulovi{\'c}}{Radulovi{\'c}}{2009}]{rad2009}
Radulovi{\'c}, D. (2009).
\newblock Another look at the disjoint blocks bootstrap.
\newblock {\em Test\/}~{\em 18\/}(1), 195--212.

\bibitem[\protect\citeauthoryear{Rho and Shao}{Rho and
  Shao}{2013}]{rho2013improving}
Rho, Y. and X.~Shao (2013).
\newblock Improving the bandwidth-free inference methods by prewhitening.
\newblock {\em Journal of Statistical Planning and Inference\/}~{\em
  143\/}(11), 1912--1922.

\bibitem[\protect\citeauthoryear{Richard}{Richard}{2009}]{richard2009modified}
Richard, P. (2009).
\newblock Modified fast double sieve bootstraps for adf tests.
\newblock {\em Computational Statistics \& Data Analysis\/}~{\em 53\/}(12),
  4490--4499.

\bibitem[\protect\citeauthoryear{van~der Vaart and Wellner}{van~der Vaart and
  Wellner}{1996}]{vanwel1996}
van~der Vaart, A.~W. and J.~A. Wellner (1996).
\newblock {\em Weak Convergence and Empirical Processes - Springer Series in
  Statistics}.
\newblock New York: Springer.

\bibitem[\protect\citeauthoryear{Volgushev and Shao}{Volgushev and
  Shao}{2014}]{volsha2014}
Volgushev, S. and X.~Shao (2014).
\newblock A general approach to the joint asymptotic analysis of statistics
  from sub-samples.
\newblock {\em Electronic Journal of Statistics\/}~{\em 8}, 390--431.

\bibitem[\protect\citeauthoryear{White}{White}{2000}]{white2000reality}
White, H. (2000).
\newblock A reality check for data snooping.
\newblock {\em Econometrica\/}~{\em 68\/}(5), 1097--1126.

\bibitem[\protect\citeauthoryear{Zhang, Shao, Hayhoe, and Wuebbles}{Zhang
  et~al.}{2011}]{zhang2011testing}
Zhang, X., X.~Shao, K.~Hayhoe, and D.~J. Wuebbles (2011).
\newblock Testing the structural stability of temporally dependent functional
  observations and application to climate projections.
\newblock {\em Electronic Journal of Statistics\/}~{\em 5}, 1765--1796.

\end{thebibliography}

\end{document}